\documentclass{article}
\usepackage[T1]{fontenc}
\usepackage[latin1]{inputenc}
\usepackage{graphics}
\usepackage{graphicx}
\usepackage{longtable}
\usepackage{color}
\usepackage{pslatex}
\usepackage{hyperref}
\usepackage{sverb}
\usepackage{subfigure}

\begin{document}
\begin{titlepage}
 
\begin{flushright} 
{\bf  IFJ PAN-IV-2016-19 
} 
\end{flushright}
 
\vskip 30 mm
\begin{center}
{\bf\huge  Potential for optimizing  Higgs boson CP measurement }\\
\vskip 5 mm
{\bf\huge  in $H \to \tau \tau$ decay   at LHC and ML techniques}\\
\end{center}
\vskip 13 mm

\begin{center}
   {\bf R. J\'ozefowicz$^{a,\star}$, E. Richter-Was$^{b}$ and Z. Was$^{c}$  }\\
   \vskip 3 mm
{\em $^a$ Open AI, San Francisco, CA, USA} \\
       {\em $^b$ Institute of Physics, Jagellonian University, Lojasiewicza 11, 30-348 Krakow, Poland} \\
       {\em $^c$ Institute of Nuclear Physics Polish Academy of Sciences , PL-31342 Krakow, Poland}\\ 
\end{center}
\vspace{1.1 cm}
\begin{center}
{\bf   ABSTRACT  }
\end{center}

We investigate potential for measuring CP state of the Higgs boson in the $H \to \tau \tau$ decay 
with consecutive  $\tau$-lepton decays in channels: $\tau^{\pm} \to \rho^{\pm} \nu_{\tau}$ 
and $\tau^{\pm} \to a_{1}^{\pm} \nu_{\tau}$ combined. Subsequent decays 
$\rho^{\pm} \to \pi^{\pm} \pi^{0}$,  $ a_{1}^{\pm} \to \rho^{0} \pi^{\pm}$ and
 $ \rho^{0}\to \pi^{+}\pi^{-}$ are taken into account. 
We will explore extensions 
of the  method, where acoplanarity angle for the planes build on the visible decay
products,   $\pi^{\pm} \pi^0$ of
 $\tau^{\pm} \to \pi^{\pm} \pi^0 \nu_{\tau}$, was used. The 
angle  is sensitive to transverse spin correlations, thus to parity.

We show, that  in the case of the cascade decays of $\tau \to a_1 \nu$, 
information on the CP state of Higgs can be extracted  
from the acoplanarity angles as well.
Because in the cascade decay 
$ a_{1}^{\pm} \to \rho^{0} \pi^{\pm},\; \rho^{0} \to \pi^+ \pi^-$ up to four planes 
can be defined,  up to 16 distinct acoplanarity angles are available for
$H \to \tau \tau \to a_1^{+} a_1^{-} \nu \nu$ decays. 
These acoplanarities carry  in part supplementary
but also correlated information. It is thus cumbersome to evaluate an overall 
sensitivity.

We  investigate sensitivity potential  of such analysis, by developing and 
implementing model in the Machine Learning (ML) techniques. We 
quantify possible improvements when  multi-dimensional
phase-space of outgoing decay products directions is used, instead of
1-dimensional projections i.e. the acoplanarity angles. 

We do not take into account ambiguities resulting from detector 
uncertainties or background contamination, we concentrate on
usefulness of ML methods and $\tau \to 3\pi \nu$ decays for Higgs boson parity 
measurement.

\vskip 1 cm


\vspace{0.2 cm}
 
\vspace{0.1 cm}
\vfill
{\small
\begin{flushleft}
{    IFJ PAN-IV-2016-19
\\ August 2016
}
\end{flushleft}
}
 
\vspace*{1mm}
\footnoterule
\noindent
{\footnotesize   {\em $^\star$ Work performed while at Google, 75 Ninth Avenue, New York, NY 10011, USA} 
}
\end{titlepage}

\section{Introduction}
The discovery of the Higgs boson by LHC experiments \cite{Aad:2012tfa,Chatrchyan:2012xdj} was followed by the measurements of its basic quantum numbers, 
such as spin~\cite{Aad:2013xqa,Khachatryan:2014kca}, mass~\cite{Aad:2015zhl}, 
couplings to fermions and bosons~\cite{Khachatryan:2016vau} and parity \cite{Khachatryan:2014kca,Aad:2015mxa}.
In particular, measurement of the CP state was possible, but so far only in the diboson decays. 
The exploration whether the discovered Higgs boson is a pure CP-even state in its coupling to fermions, 
or if admixture of the odd state is present, is of prime interest for the LHC physics program of
the next years. 
If the CP-odd component is detected it would be an evidence of the new physics: 
the non-standard CP-violation. One way to probe parity of the $H$ couplings to fermions
is discussed in the literature since decades \cite{Kramer:1993jn}.  
It relies on the measurement of  transverse spin correlations in the $H \to \tau \tau$ decay.
 
The loss of neutrinos for detection, and detector precision
was limiting the original method of~\cite{Kramer:1993jn}. The possible way out,  was found in \cite{Bower:2002zx}. 
It relied on measurement of spin correlations between visible decay product of $\tau$'s, more precisely the 
$\pi^\pm \pi^0$ from the dominant $\tau$ decay channel: $\tau^\pm \to \pi^\pm \pi^0 \nu_\tau$.
This decay channel of single intermediate spin one state is
saturated by $\rho^\pm$. It was relatively easy to define corresponding observable: 
acoplanarity angle of two planes (defined in $\rho^+ \rho^-$ pair rest frame),  spanned respectively
on $\pi^+ \pi^0$ and  $\pi^- \pi^0$. To achieve sensitivity to CP state, events had to be split into two groups;
relative sign of energy differences for $\pi^\pm$ and $\pi^0$ for each of the two pairs was used 
to separate events.
The method presented  in \cite{Bower:2002zx}, does not need much adaptation for 
the  LHC measurements, see eg. \cite{Przedzinski:2014pla}.
Similarly as in the Linear Collider case,
there is no need to reconstruct   $\tau$ lepton rest-frames, it was enough to reconstruct rest frame  
of the $\rho^{\pm}\  \rho^{\mp}$ pair. 

A prospect to extend measurements to other $\tau$ decay modes is tempting.
The $\tau$ lepton decays  into  variety of channels. More
than  20 different decay channels have been observed.
In the present paper we will study extension of the acoplanarity angle  method to other sizable $\tau$ decay channel, namely 
 $\tau^\pm \to (3\pi)^\pm \nu$.

Started in Ref.~\cite{Rouge:2005iy}  effort to exploit all 1-prong $\tau$ 
decay modes lead to several publications
 \cite{Berge:2008dr,Berge:2011ij,Berge:2013jra,Berge:2014sra,Berge:2015nua}.
 An idea that position of decay $\tau$ vertex could be 
used as a backbone of the measurement was promising substantial gain in 
sensitivity, certainly better than if vertex reconstruction
was assumed to be at the measurability limit only \cite{Desch:2003mw}. Use of the 
decay vertex is expected to be particularly promising in case of 3-prong $\tau$ decays, with already achieved performance of tracking and vertexing of LHC experiments  \cite{Aad:2015unr,Chatrchyan:2012zz,Aaij:2014jba}.    

In the following study, let us concentrate, on measurement  sensitivity 
which can be achieved for $H\to \tau \tau$ data without
exploitation of  $\tau$ decay vertex.
Instead we will explore in a greater detail effects due to spin correlations
in cascade $\tau$ decays.
In Table~\ref{tab:tauBR} we show branching ratios for specific modes discussed in this analysis. Inclusion of 
$ a_{1}^{\pm} \to \rho^{0} \pi^{\pm},\ \rho^{0} \to  \pi^{+} \pi^{-} $ or  $ a_{1}^{\pm} \to 2 \pi^{0} \ \pi^{\pm} $  increases
the available fraction of $H \to \tau\tau$ decay rate from 6.5 to  19.2 \%.
For the analysis discussed here we will consider either $\tau$ decay to 3 charged $\pi$
 or $\tau$ decay to $\pi^{\pm} \ \pi^{0}$ with 
intermediate  $\rho^{\pm}$ resonance. That results in an increase of statistical sample by a factor of nearly two,
from 6.5  to 11.9 \%.

Our paper is organized as follows. In Section 2 we demonstrate that indeed acoplanarity angles can be defined in case 
of $\tau^{\pm} \to a_{1}^{\pm} \nu$ decays and if events are split into appropriate groups, sensitivity to CP parity of 
the Higgs boson can be achieved for this 1-dimensional variables. 
Because several such angles can be defined for the $\tau$ decays into more than two scalars, we demonstrate in Section 3, that all resulting observables, if combined together
using methods of Machine Learning (ML), like Deep Learning Neural Network, can improve sensitivity. 
We discuss also, how information has to be prepared for Neural Network to assure sensitivity. Summary, Section 4, closes the paper. 

\section{ The 1-dimensional analyzes} \label{sec:physics}

Let us discuss now in more details observables for the three cases:
(i) $ H \to \tau\ \tau \to  \rho^{\pm}   \ \rho^{\mp} \ 2\ \nu  $,
(ii) $  H \to\tau\ \tau \to  \rho^{\pm}   \ a_{1}^{\mp} \ 2\ \nu  $ and (iii)  $  H \to \tau\ \tau \to  a_{1}^{\pm}   \ a_{1}^{\mp} \ 2\ \nu  $.
The rates for  those configurations with respect
to the total $H \to \tau \tau$ rate are  6.5\%, 4.6\% and 0.8\% respectively,
see Table~\ref{tab:tauBR}. 
The relatively well established
strategy for CP measurement exists for  the configuration (i).  
One can  significantly increase available for analysis statistics if  also (ii) and (iii) 
configurations are used. At this moment, we have no intention to explore
a $\pi^0\pi^0\pi^\pm$ tau decay, even though from the physics point of view it would be straightforward and available sample of 
events would again increase significantly, see Table~\ref{tab:tauBR}. 
This case, because of experimental context requires  evaluation if 2$\pi^0$'s can be resolved.

The  {\it 1-dimensional analysis}, as discussed in this section, 
will rely on building 1-dimensional observable, or a few of them,
sensitive to the CP nature of the Higgs boson.
Below, we propose those observables, but without  an attempt to quantify 
possible sensitivity of the experimental analysis, especially consequences of existing correlations between variables.

We   extend  definition of an acoplanarity angle introduced for  
the  $\tau \to \pi^{\pm} \pi^0 \nu$ decay to a few more  available for  $\tau^{\pm} \to a_{1}^{\pm} \nu$  decay
and extend also definition of variable which can be used to split events into groups necessary for discrimination
between different CP states.
The subtle effect of the  CP-even vs  CP-odd nature  manifests itself in 
these angular distributions of $\tau$ decay products: 
the acoplanarity angle(s) $\varphi^*_{i,k}$ between  planes defined by  the visible decay products in   
the hadronic cascades starting from  the $\tau$ leptons. 
The sub-scrips $i,k$ are used to index elementary decays of the cascades and
 the corresponding decay planes. Introduction of indices is convenient  
if there are more than two visible decay products of the $\tau^\pm$.

For numerical studies, we use Monte Carlo events of the SM, 125 GeV mass, 
Higgs boson, produced in pp collision at 13~TeV centre-of-mass energy,
generated with {\tt Pythia 8.2} and with spin correlations simulated using {\tt TauSpinner} \cite{Przedzinski:2014pla} package.
For $\tau$ lepton decays we use {\tt Tauolapp} library \cite{Davidson:2010rw}. All spin and
parity effects are implemented with the help of {\tt TauSpinner}  weight $wt^{CP}$. 
That is why, samples prepared for CP even or odd Higgs are correlated.

In  our studies, to emulate partly detector conditions, 
a minimal set of cuts is used. 
We require that the transverse momenta of the visible decay products combined, for each $\tau$, is bigger than 20 GeV. 
It is also required that transverse momentum of each $\pi^\pm$ is bigger than 1 GeV.

\subsection{ The $ H \to \tau\ \tau \to  \rho^{\pm}   \ \rho^{\mp} \ 2\ \nu  $ case}

This final state configuration contributes about 6.5\% of all $H \to \tau\ \tau$ decays. 
The method of using the acoplanarity $\varphi^*_{\rho^+ \rho^-}$ of the $\rho^{\pm}$ decay planes  in
$\rho^{\pm} - \rho^{\mp}$ centre-off-mass system
 as sensitive observable, proposed in~\cite{Bower:2002zx},
has been so far considered as the most promising one. 
This method requires that the track from the charged $\pi^{\pm}$ and a cluster from $\pi^{0}$ can be separated,
which is now well within experimental reach of the LHC detectors~\cite{Aad:2015unr,Chatrchyan:2012zz}. 

The sample  is split into two sub-groups, with the  help of  the sign for the  
differences of $\pi^{\pm}$ and $\pi^{0}$ energies.
For each $\tau^\pm$ in the pair, we define 
\begin{equation}
y_\rho^{\pm} = \frac{ E^{\pi^{\pm}} - E^{\pi^{0}}}{ E^{\pi^{\pm}} + E^{\pi^{0}}} 
\label{Eq:yrho}
\end{equation}
and we group events into two categories depending on the sign of the product $y_\rho^+ y_\rho^-$. 
The $E^{\pi^{\pm}}$  and $E^{\pi^{0}}$ 
can  be defined in the $\rho^{+} \rho^{-}$ rest frame
or in laboratory frame, with no sizable difference of the sensitivity.  
In each group, acoplanarity angle $\varphi^*_{\rho^+, \rho^-}$ shows 
nice sinusoidal shape which is shifted by $180^\circ$ 
between the cases of  scalar and pseudo-scalar Higgs. 
Note that if
the shift was present but was smaller than $180^\circ$, the size of 
that shift would measure 
the mixing between  two CP states of the Higgs boson. 
We will return to this point in Subsection \ref{sec:CPmix}.
Fig.~\ref{fig:rhorho} shows overlaid distributions for 
of the scalar (black) and pseudo-scalar (red) models, for each group of events separated by the sign of the 
product  $y_\rho^+ y_\rho^-$.  On the Fig.~\ref{fig:rhorho} we show also
line shape of the  $\pi^\pm \pi^0$  mass, for system
 originating from the same $\tau$.

\subsection{ The $ H \to \tau\ \tau \to  a_{1}^{\pm} \ \rho^{\mp} \ 2\ \nu  $ case}

\begin{table}
  \begin{center}
  \begin{tabular}{|l|l|r|r|c|}
  \hline\hline 
  Decay mode                      & Cascade decay                                                 & $\tau$ BR. & Cumul. $H\to\tau\tau$ frac.    & Used for analysis    \\ 
  \hline\hline
  $ \tau^{\pm} \to \rho^{\pm} \nu$  & $ \rho^{\pm} \to \pi^{0}\ \pi^{\mp} $                                    &  25.5\%        &   6.5\%     & Yes                   \\ 
  \hline
  $ \tau^{\pm} \to a_{1}^{\pm} \nu$  & $ a_{1}^{\pm} \to \rho^{0} \pi^{\mp},\ \rho^{0} \to  \pi^{+} \pi^{-} $    &  9.0\%          &  11.9\%     & Yes                  \\ 
                                  & $ a_{1}^{\pm} \to 2 \pi^{0} \ \pi^{\mp} $                                &  9.3\%         &  19.2\%      & No                   \\ 
  \hline\hline
\end{tabular}
\end{center}
\caption{ Branching ratios of the $\tau$ lepton decay modes \cite{Agashe:2014kda}, 
and resulting, cumulated fraction of $H\to \tau\tau$ events available for 
parity analysis. }
\label{tab:tauBR}
\end{table}
\begin{figure}
  \begin{center}                               
{
   \includegraphics[width=7.5cm,angle=0]{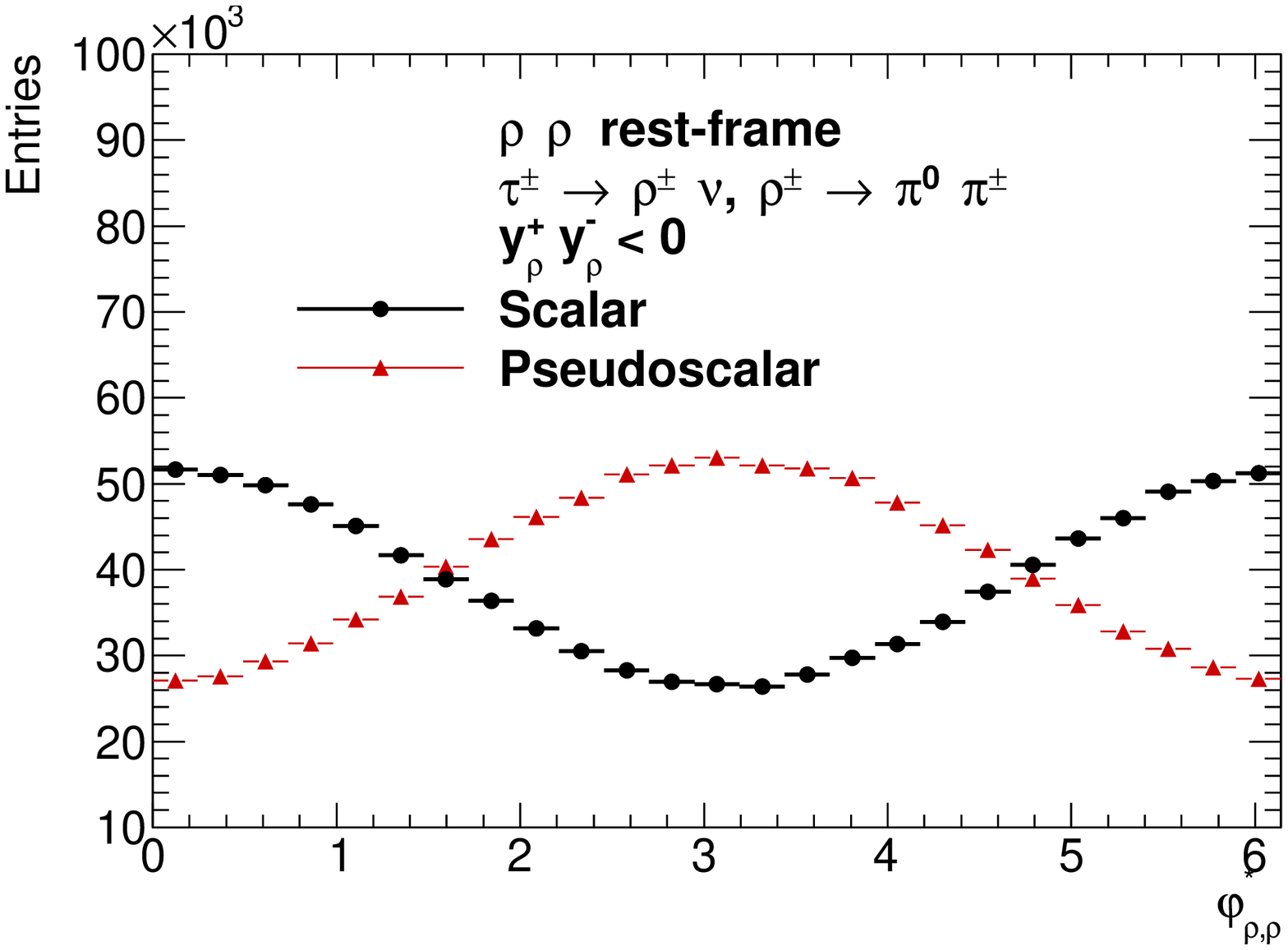}
   \includegraphics[width=7.5cm,angle=0]{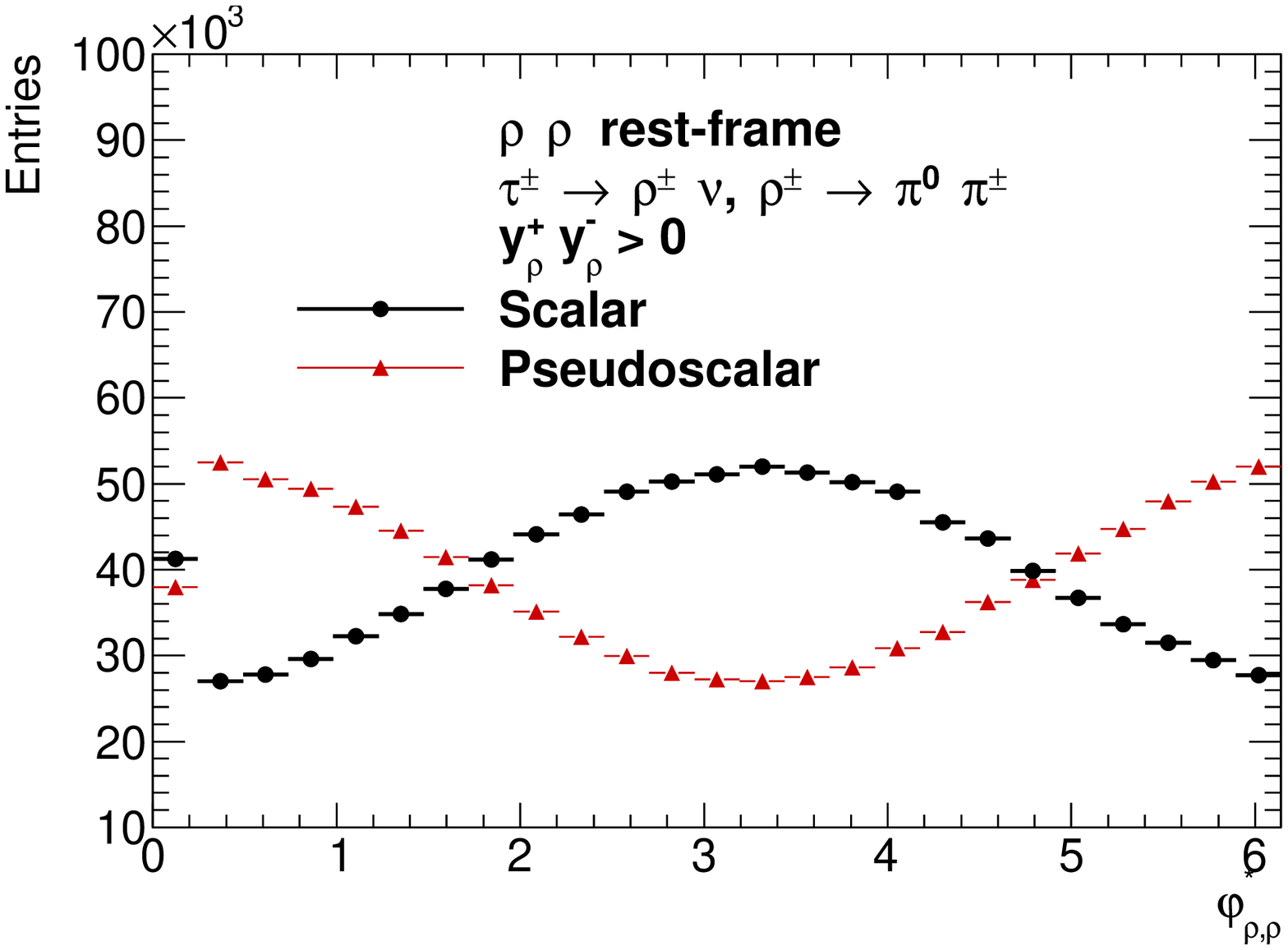}
   \includegraphics[width=7.5cm,angle=0]{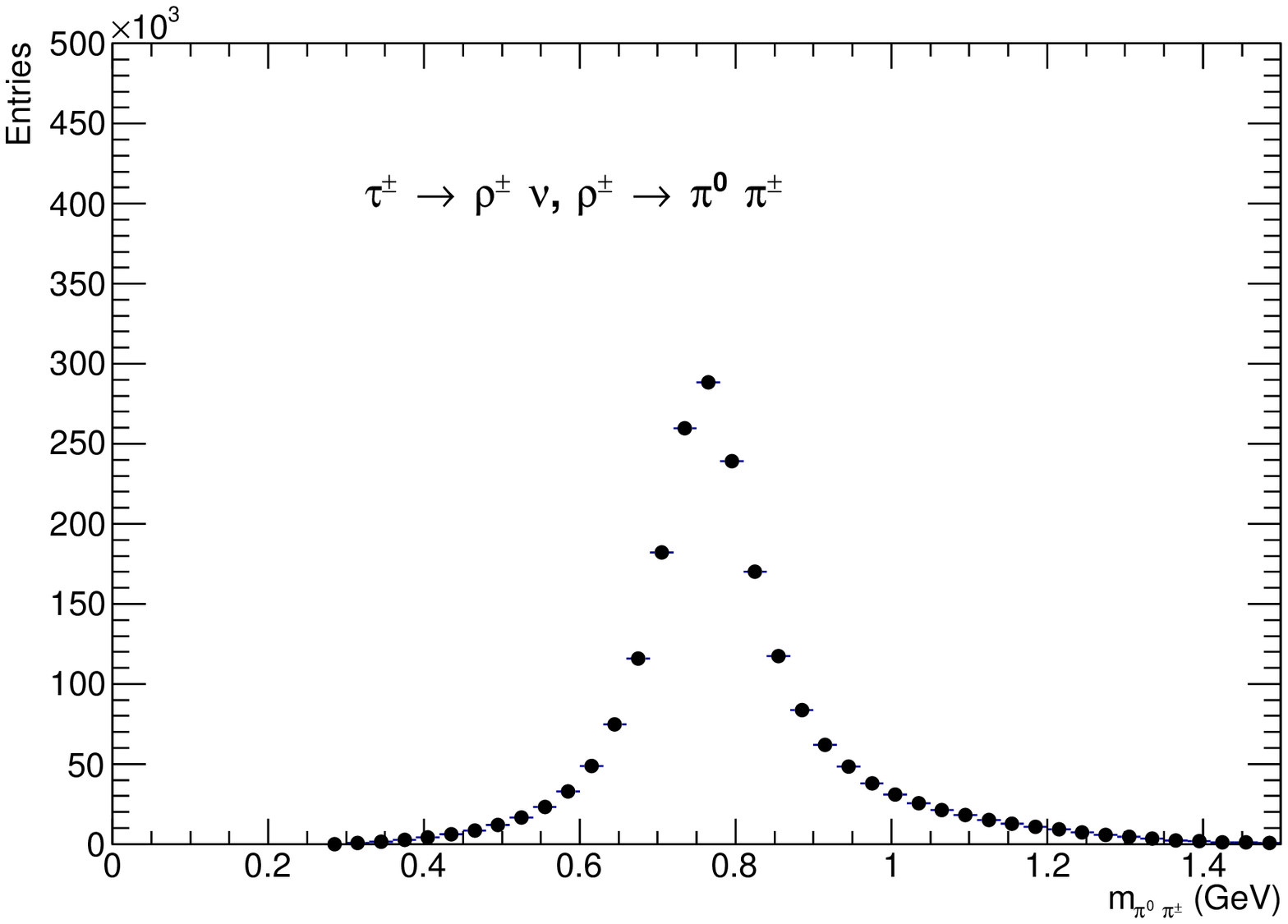}
}
\end{center}
\caption{ Acoplanarity angle $\varphi^*_{\rho \rho}$ of oriented half decay plane  
for the decay products of $\rho^+$ and $\rho^-$.
Events are grouped by the sign of $y_{\rho}^{+} y_{\rho}^{-}$.
Invariant mass of  the $\rho^{\pm} \to \pi^{+} \pi^{-}$ system (bottom) is also shown. 
\label{fig:rhorho} }
\end{figure}

This final state configuration contribute about 4.6\% of all $H \to \tau\ \tau$ decays. 
Let us  extend  the  method explained in the previous sub-section to a new case,
where
there are two possibilities for the $a_1$ system decay plane definition. 
One can take the  
decay products of primary resonance
$ a_{1}^{\pm}$, or of secondary resonance $ \rho^{0}$ originating from $\tau^\pm$ decays. 
Also, in case of 
$ a_{1}^{\pm} \to 3 \pi^{\pm}$ decay, we do not know which combinations of $\pi^{+} \pi^{-}$ form the intermediate $\rho^{0}$ resonance, 
for each event we take into account both combinations. Counting possible pairing,
we finally construct 4 planes, thus 4 distinct acoplanarities with the plane of
$\rho^\mp$, from the other $\tau$, decay. 
We could use reconstructed invariant masses, to decide which of the two
$\pi^{+} \pi^{-}$ pairs is closer 
to the $\rho^0$ peak, and choose the corresponding acoplanarities only,
but it was not straightforward to evaluate the impact on optimalization.

As in the previous case of Fig.~\ref{fig:rhorho},  the discriminating shapes of the acoplanarity angles between scalar and pseudo-scalar CP states 
appear  only if the samples are split into sub categories. In case of the $\tau^{\pm} \to a_{1}^{\pm} \nu $ decay we calculate 
$y_{a_1}^{\pm}$ and $y_{\rho^0}^{\pm}$

\begin{equation}
y_{\rho^{0}}^\pm = \frac{ E^{\pi^{+}} - E^{\pi^{-}}}{ E^{\pi^{+}} + E^{\pi^{-}}}, \ \ \ \  
y_{a_1}^{\pm} = \frac{ E^{\rho^{0}}-E^{\pi^{\pm}}}{ E^{\rho^{0}} + E^{\pi^{\pm}}} - 
 \frac{m^2_{a_1}-m^2_{\pi^{\pm}}+m^2_{\rho^0}}{ 2 m^2_{a_1}}
\label{Eq:ya1}
\end{equation}

In this case we take into account in $y_{a_1}^{\pm}$ definition,
that masses of $\rho^0$ and $\pi^\pm$ are substantially different. The sign
of the product $ y_{a_1}^{\pm} y_{\rho}^{\mp}$  or respectively  $ y_{\rho^{0}}^{\pm} y_{\rho}^{\mp}$ is used to split events into two separate categories, 
which are almost equally populated. The $m_{a_1}$  and $m_{\rho^0}$ denote respectively the invariant mass of $(3\pi)^\pm$ system and $(2\pi)$ system which we assume 
to form $a_1$ and $\rho^0$.

\begin{figure}
  \begin{center}                               
{
   \includegraphics[width=7.5cm,angle=0]{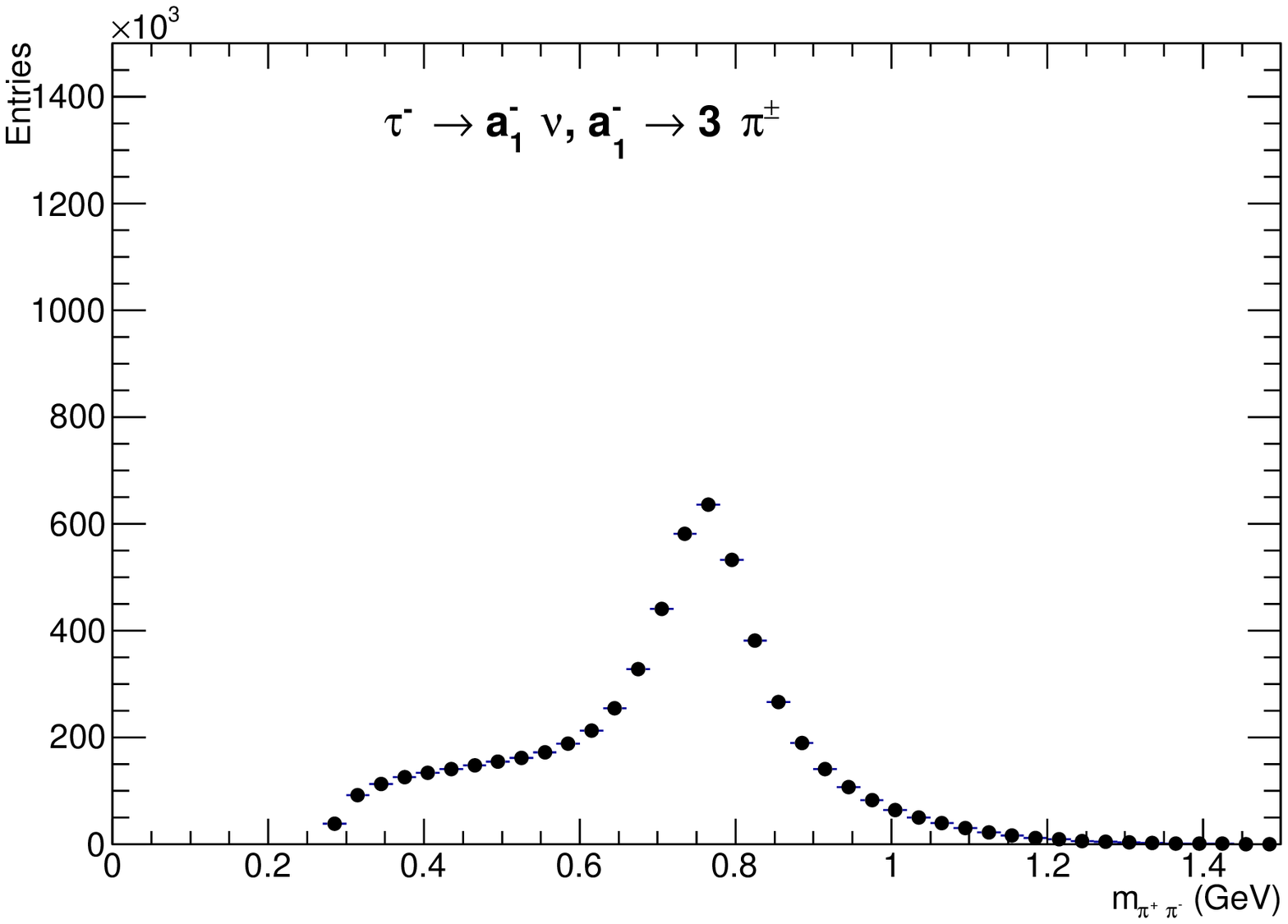}
   \includegraphics[width=7.5cm,angle=0]{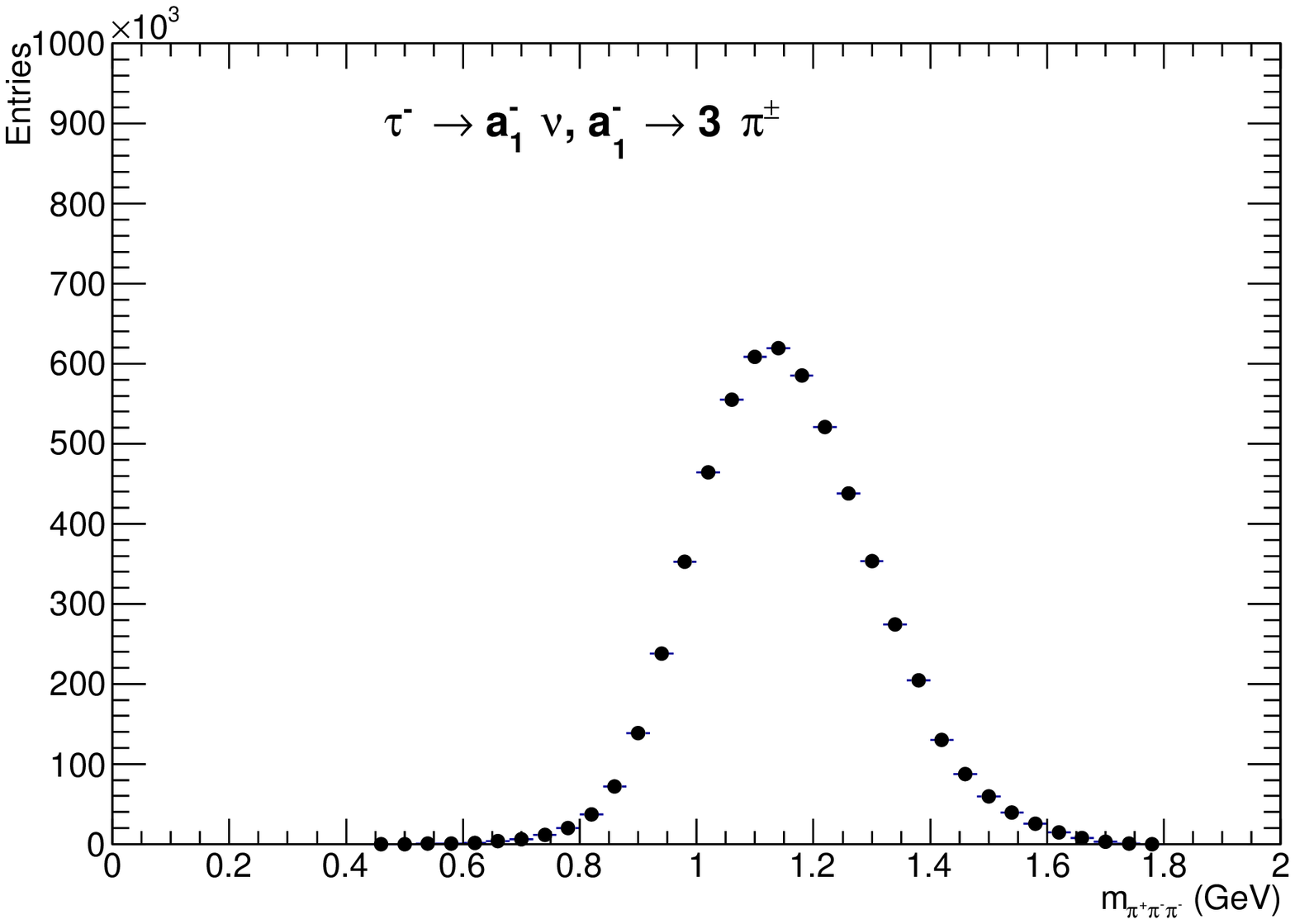}
}
\end{center}
\caption{Invariant mass of the $(3\pi)^{\pm}$ and  $\pi^{+}\pi^{-} $ systems in case  of one 
$\tau^{\pm} \to a_{1}^{\pm} \nu \to (3 \pi)^{\pm} \nu$ decays.
\label{} }
  \begin{center}                               
{
   \includegraphics[width=7.5cm,angle=0]{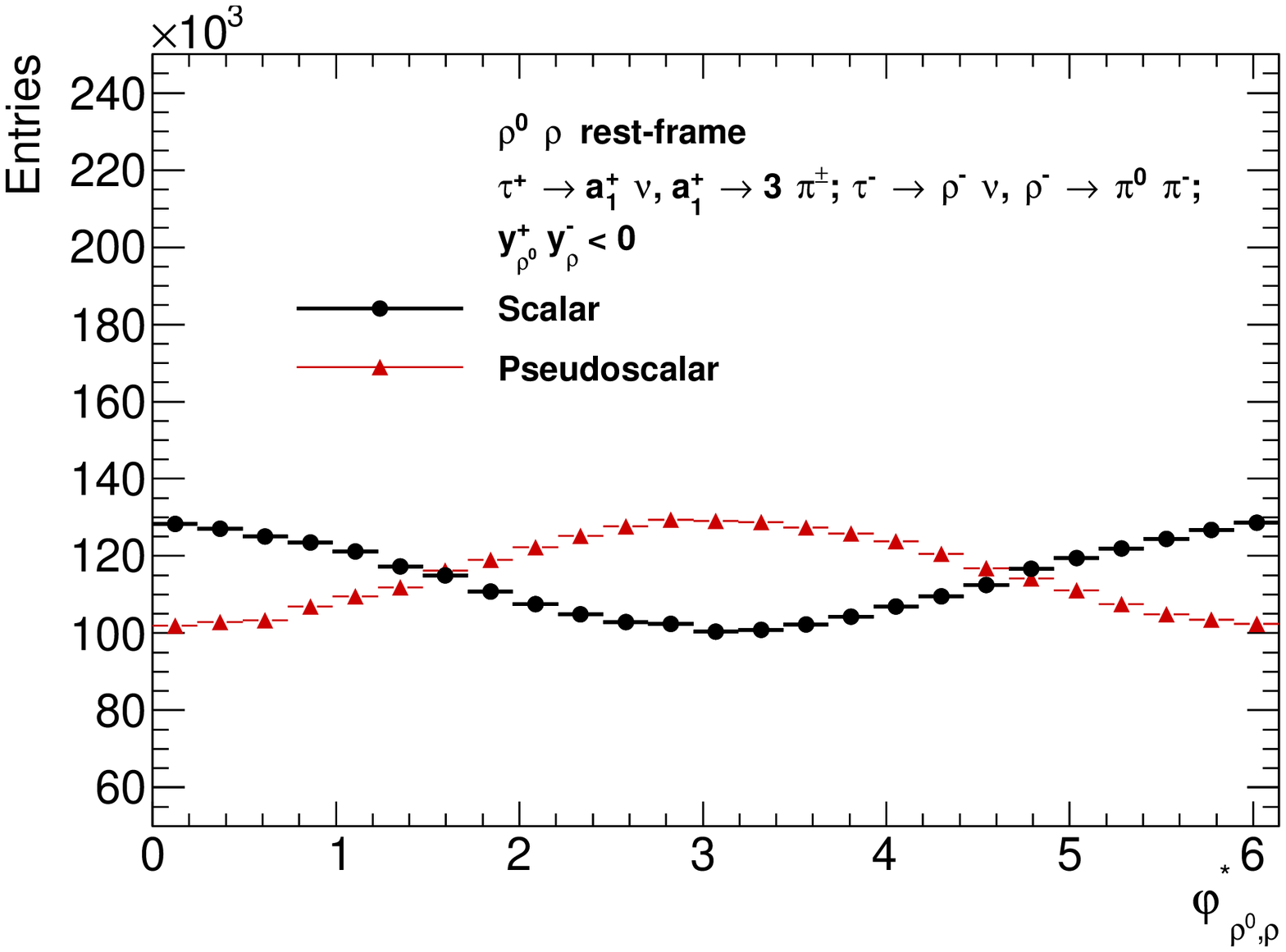}
   \includegraphics[width=7.5cm,angle=0]{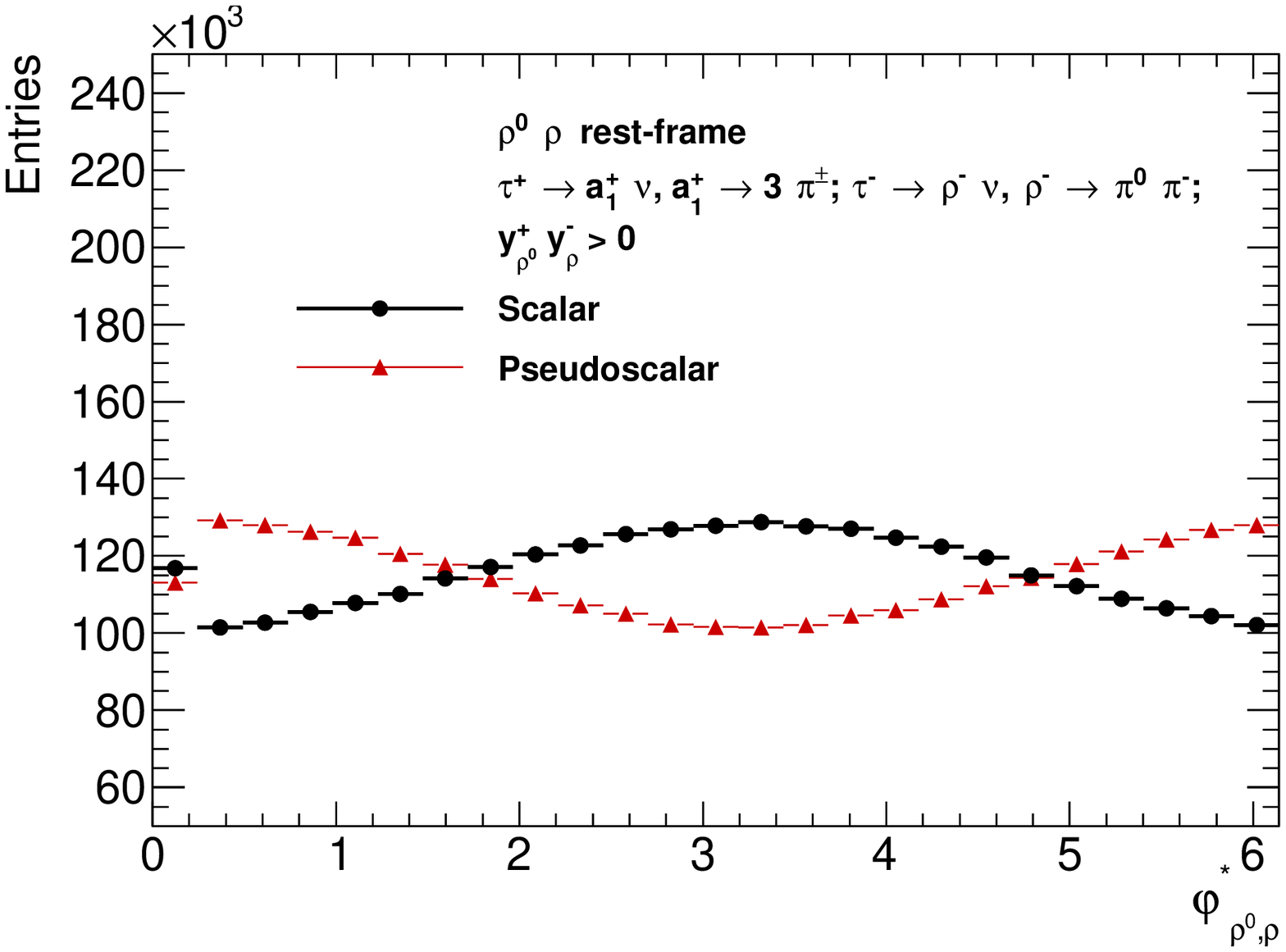}
   \includegraphics[width=7.5cm,angle=0]{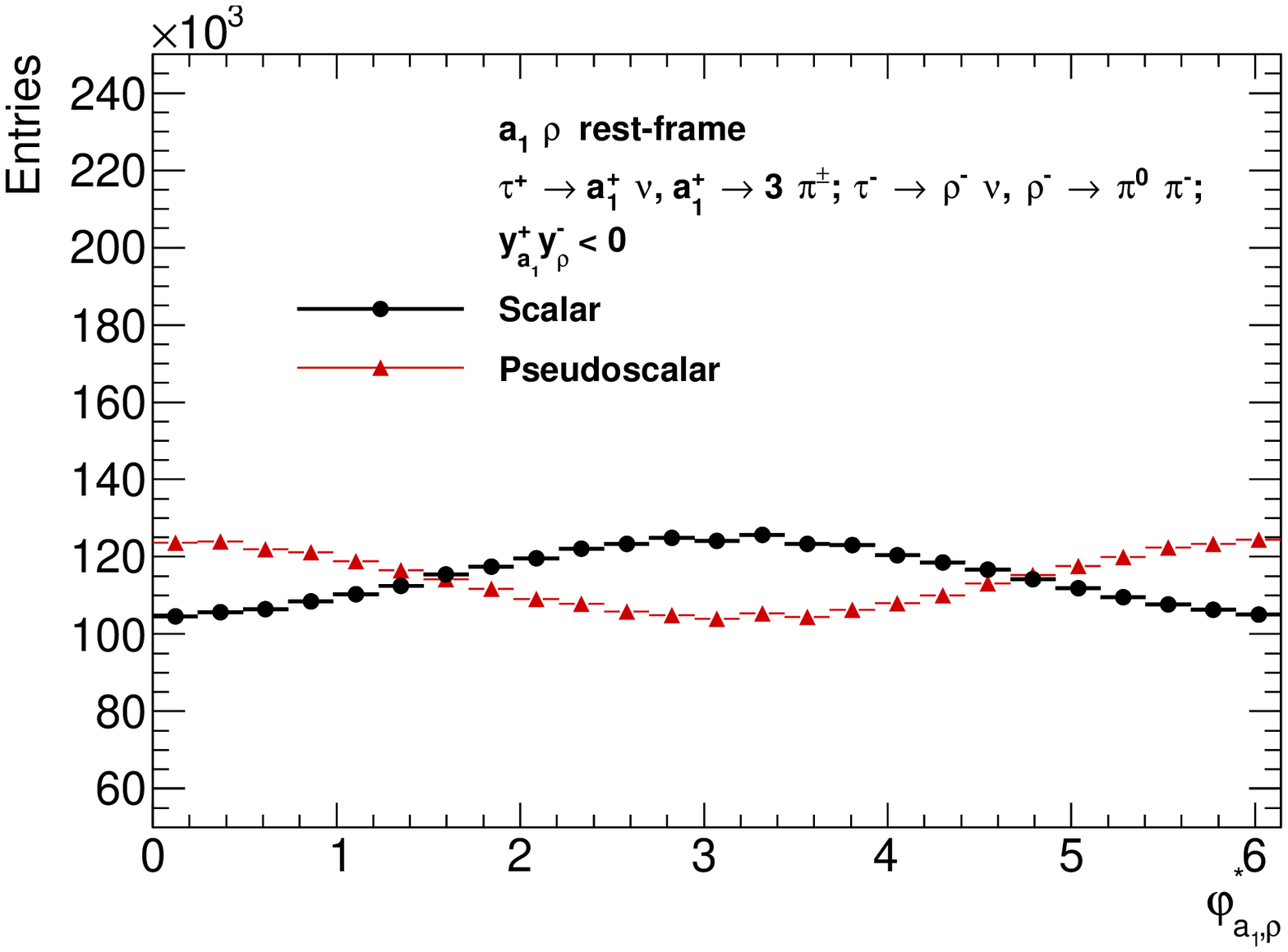}
   \includegraphics[width=7.5cm,angle=0]{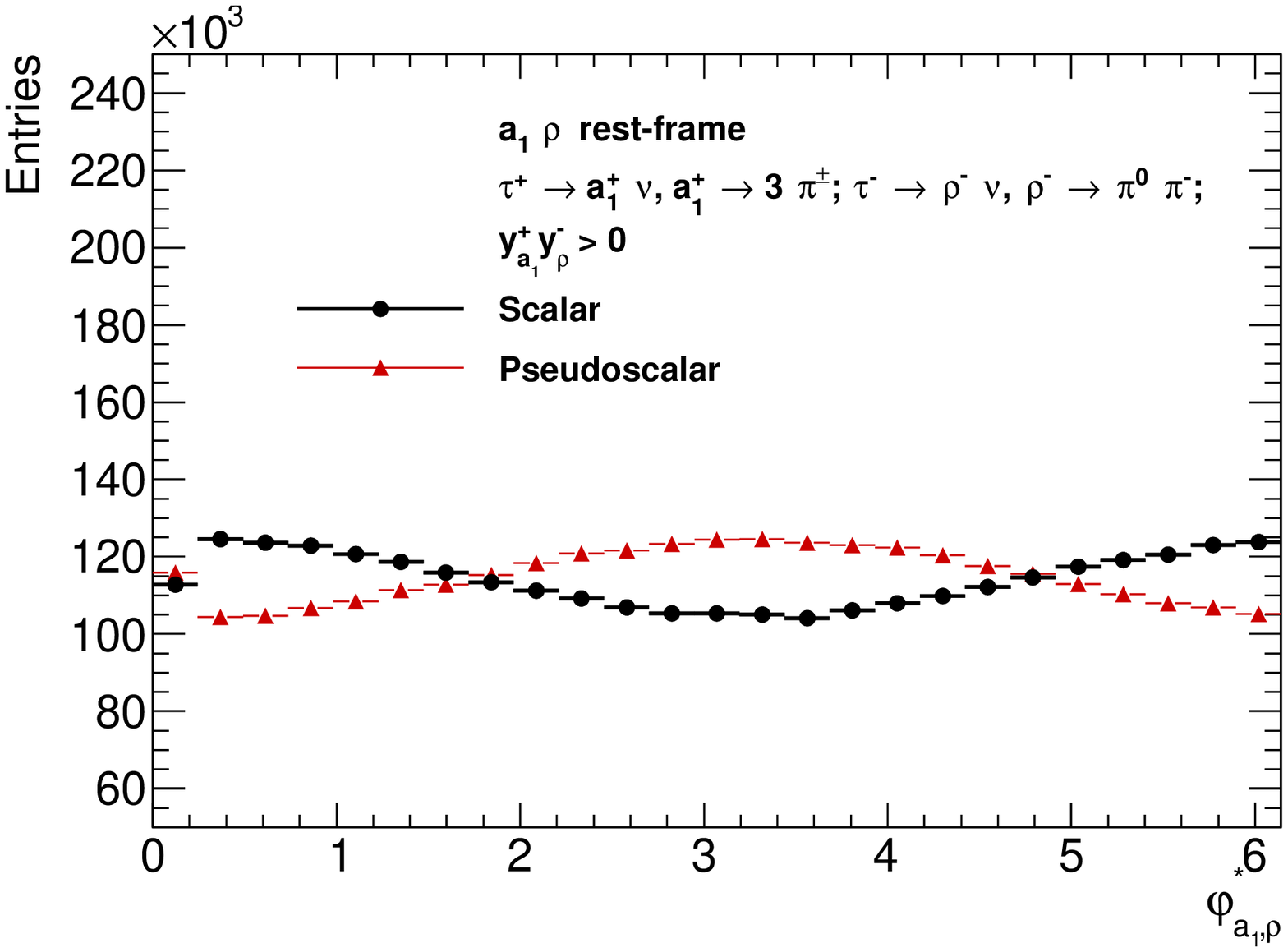}
}
\end{center}
\caption{Acoplanarity angles of oriented half decay planes: $\varphi^*_{\rho^{0} \rho}$ (top), $\varphi^*_{{a_1} \rho}$ (bottom), 
for events grouped by the sign of  $y_{\rho^{0}}^+ y_{\rho}^-$ and $y_{a_1}^{+} y_{\rho}^{-}$ respectively.
\label{fig:a1rho}}
\end{figure}

Fig.~\ref{fig:a1rho} shows overlaid shapes of the scalar (black) and pseudo-scalar (red) models. For each acoplanarity angle
events are split into two groups, to make this observable CP  sensitive. 
The amplitude of modulation is smaller than in previously discussed case, as the 
information on the CP state is now dissipated in the cascade decay of one $\tau$ lepton. 
Also it is smeared due to the possible wrong pairing for  the  $\pi^{+} \pi^{-}$ to form  
$\rho^{0}$ resonance. To illustrate effect of this wrong pairing, an invariant masses of the  $\pi^{+} \pi^{-}$ and  $\pi^{+} \pi^{-} \pi^{+}$ systems
are also shown in Fig.~\ref{fig:a1rho}. 
In total, four acoplanarity angles can be constructed. They all have quite similar sinusoidal shapes
and carry CP information. They are correlated and their CP distinctive powers are not independent.
That is why we do not quantify sensitivity in the scope of 1-dimensional 
histograms-observables. It was important  to demonstrate that sensitivity is 
indeed present, and that acoplanarity angles are again good candidate for 
the definition of 1-dimensional observables. Note that for some  plots
 of $\varphi^*_{i,k}$
it looks as if scalar/pseudo-scalar contributions
 were accidentally interchanged, it is the consequence 
of more complex nature of the 3$\pi$ decays.

\subsection{ The $ H \to \tau\ \tau \to  a_{1}^{\pm}   \ a_{1}^{\mp} \ 2\ \nu  $ case}

This final state  contribute about 0.8\% of all $H \to \tau\tau$ decays. 
We extend here the previously explained method
to case with even more combinations of decay planes: $(  a_{1}^{\pm}, a_{1}^{\mp})$, 
$(  a_{1}^{\pm}, \rho^{0})$, $(   \rho^{0}, a_{1}^{\mp})$ and $( \rho^{0}, \rho^{0})$.
In total
16 distinct acoplanarity angles can be defined. 
For each acoplanarity angle, separation into two categories, depending on the sign 
of $y_{a_1}^{\pm} y_{a_1}^{\mp}$,  $y_{a_1}^{\pm} y_{\rho^{0}}^{\mp}$, $y_{\rho^{0}}^{\pm} y_{a_1}^{\mp} $ and $y_{\rho^{0}}^{\pm} y_{\rho^{0}}^{\mp}$ products respectively is performed. Those variables are calculated as 
in Eqs. (\ref{Eq:yrho}) and (\ref{Eq:ya1}). 
In this case also, it was possible to achieve almost equal population of events separated by 
the sign of $y^+y^-$, 
and for each of 16 acoplanarity angles. 
  
In the case of $ a_{1}^{\pm} \to 3 \pi^{\pm}$, a priori, we do not know which combinations 
of $\pi^{+} \pi^{-}$ formed the $\rho^{0}$ resonance: we take therefore 
into account all possible
pairing of $\pi^{+} \pi^{-}$ in each $\tau$ decay. These four  invariant masses respectively of two possible pairs of 
$\pi^+ \pi^-$ forming $\rho^0$ defined for each of the two $\tau$'s, can be  helpful again for sensitivity optimalization.

\begin{figure}
  \begin{center}                               
{
   \includegraphics[width=7.5cm,angle=0]{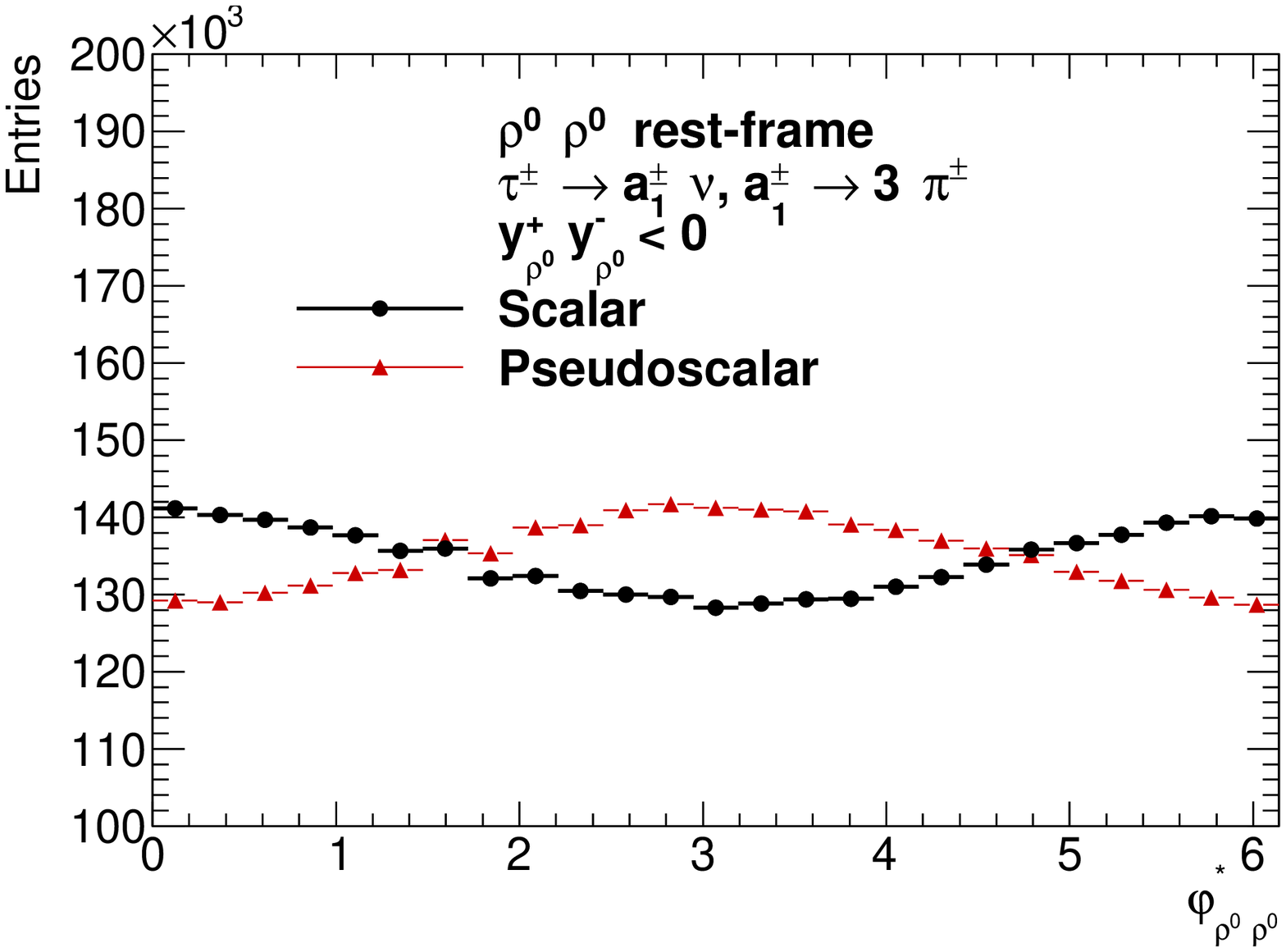}
   \includegraphics[width=7.5cm,angle=0]{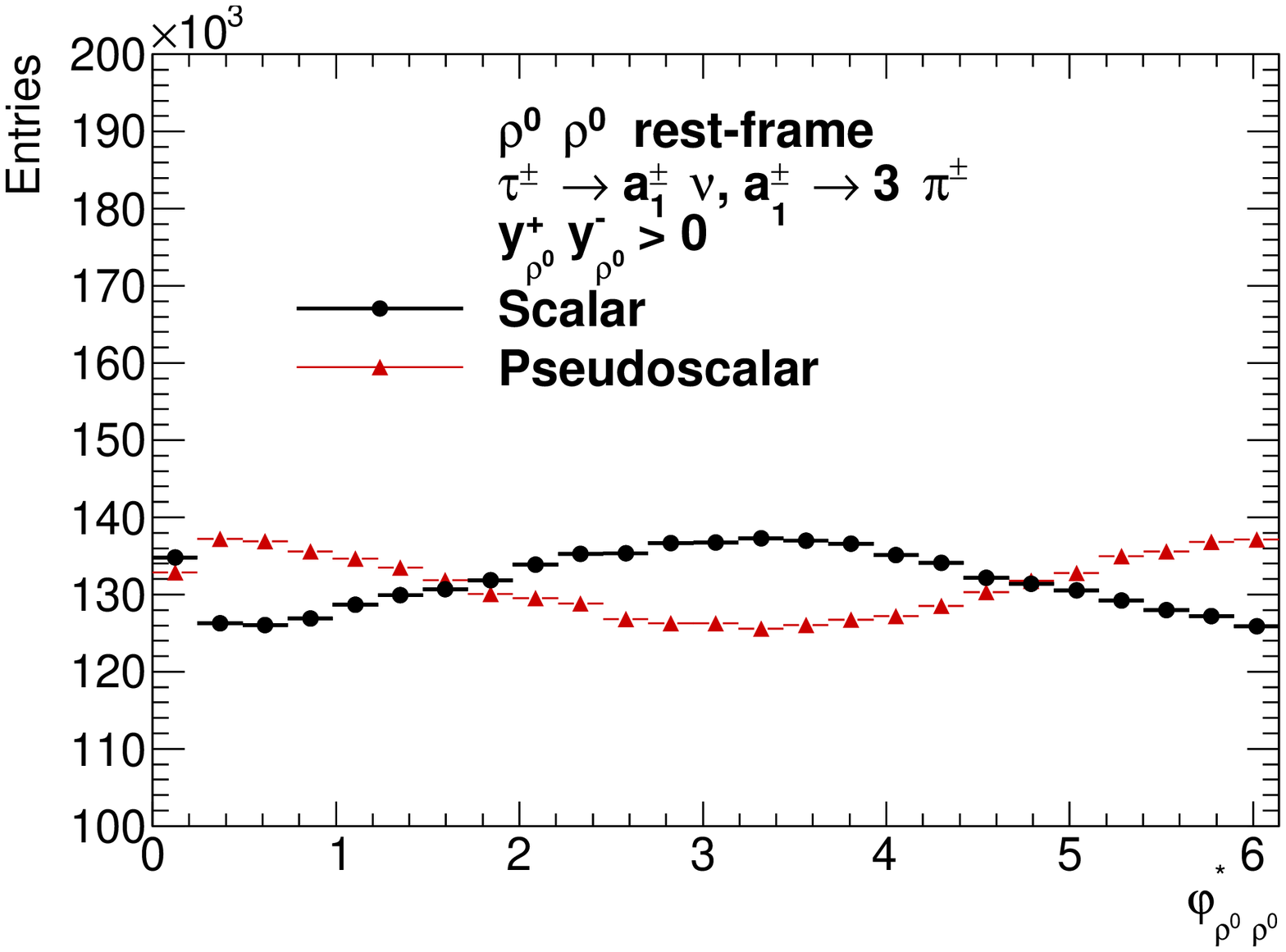}
   \includegraphics[width=7.5cm,angle=0]{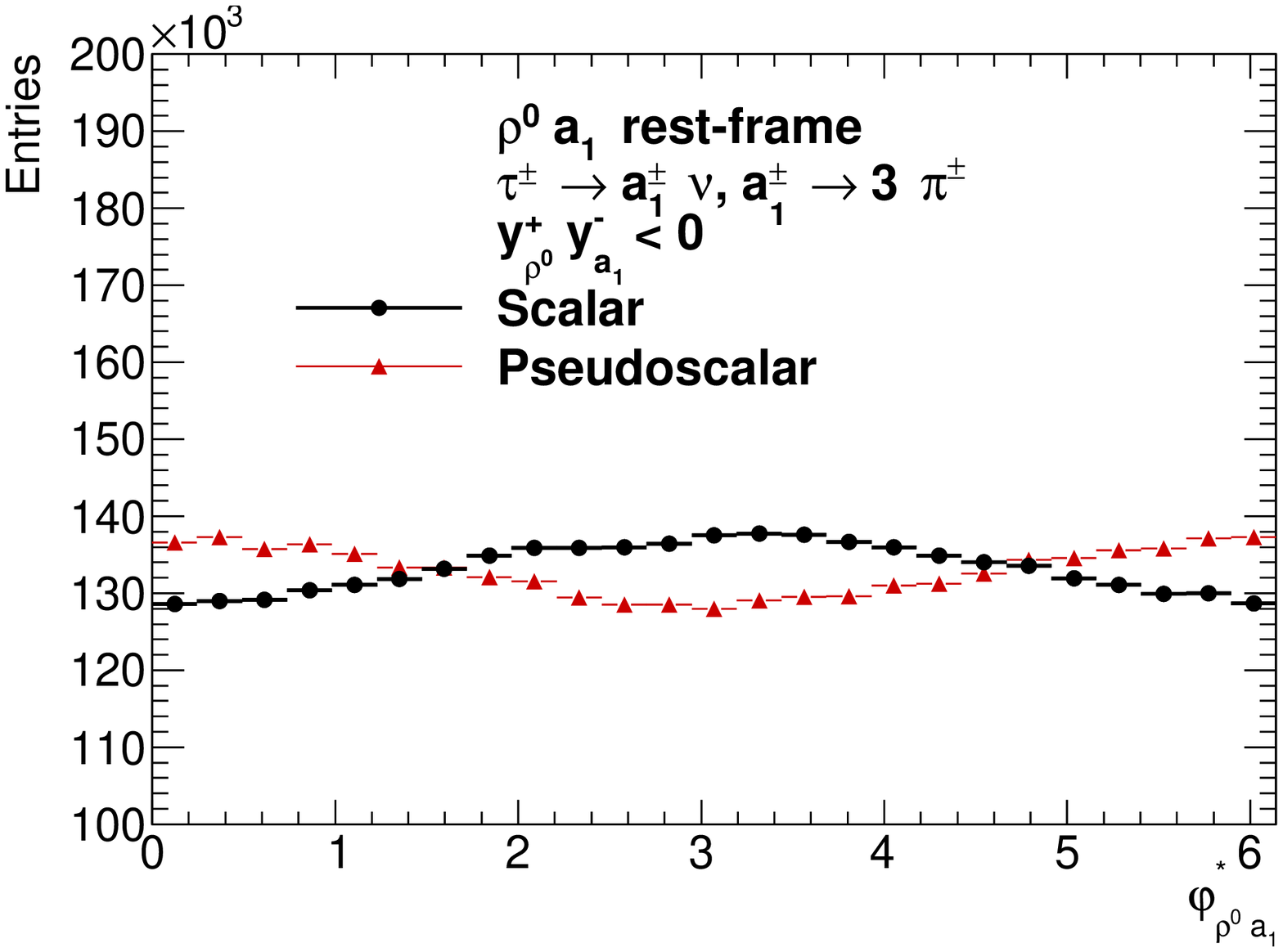}
   \includegraphics[width=7.5cm,angle=0]{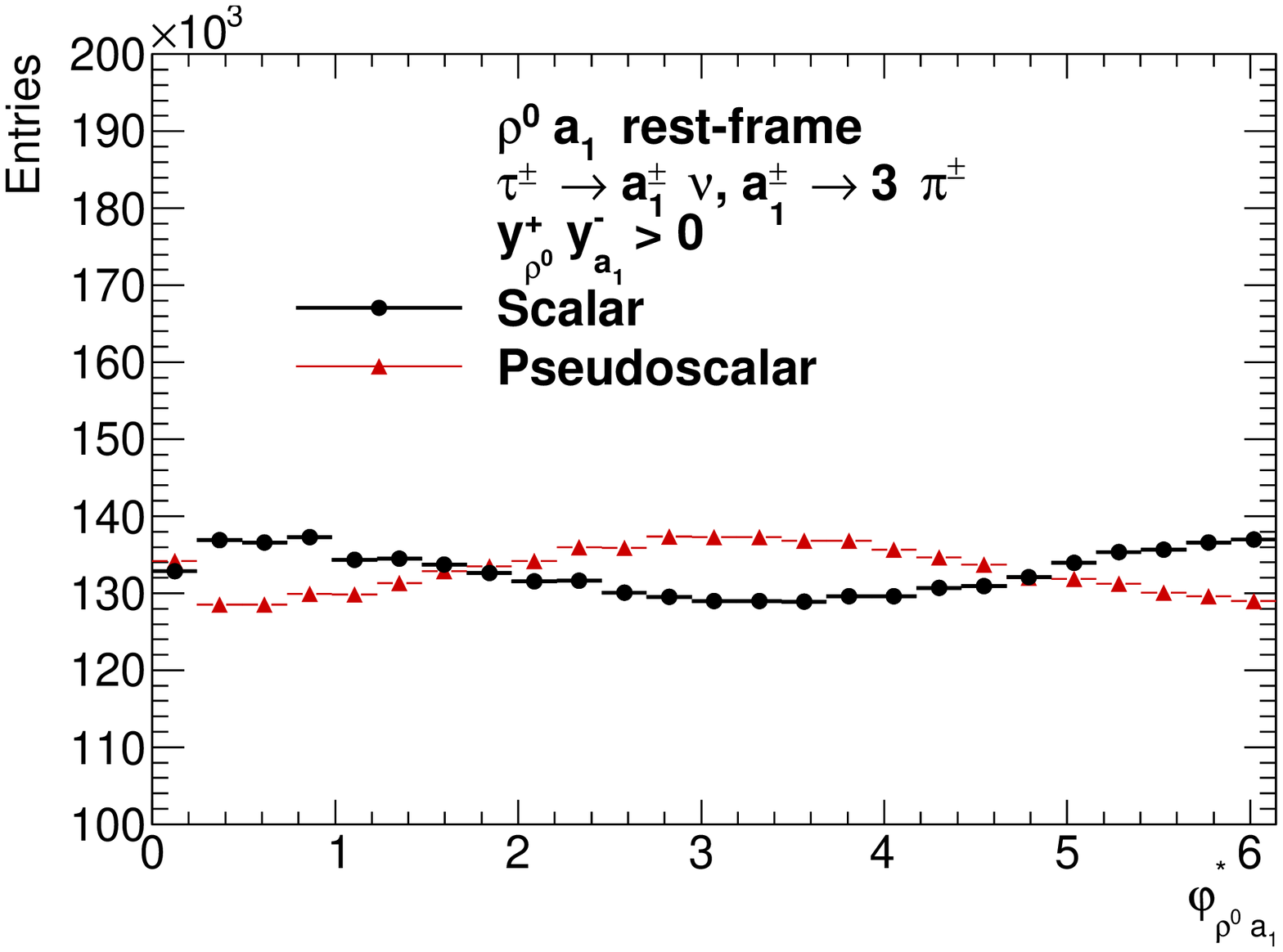}
   \includegraphics[width=7.5cm,angle=0]{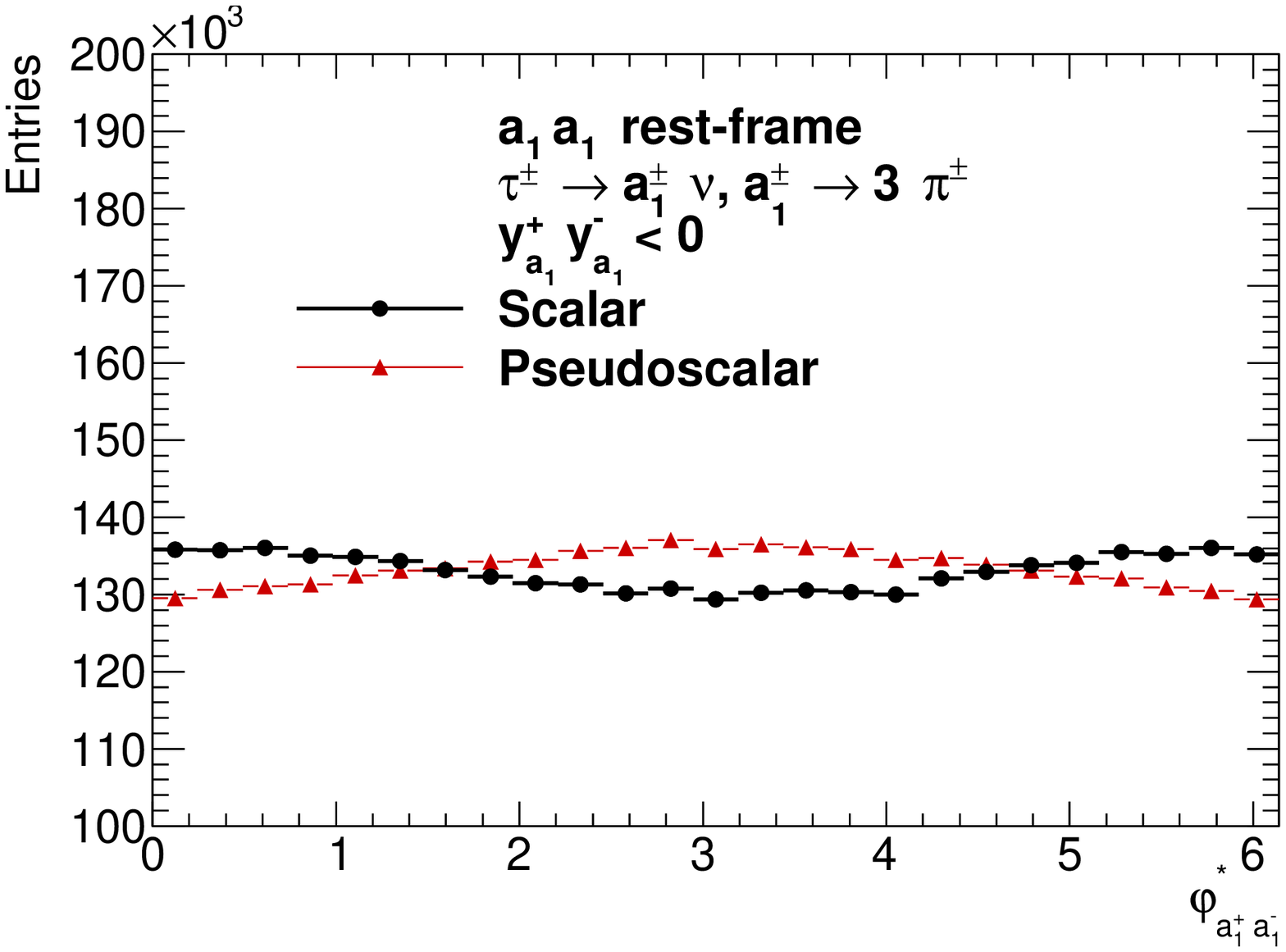}
   \includegraphics[width=7.5cm,angle=0]{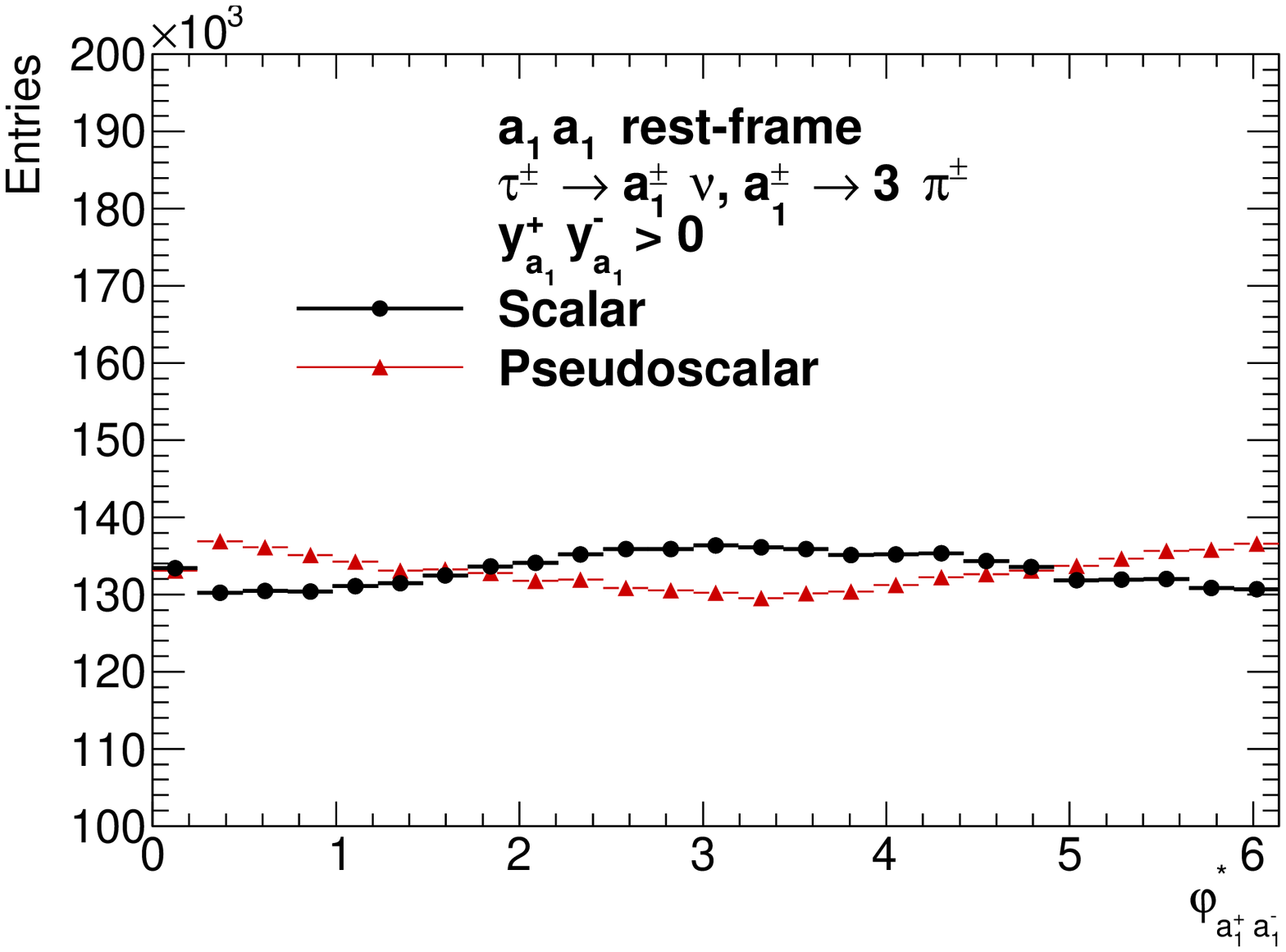}
}
\end{center}
\caption{Acoplanarity angles of oriented half decay planes: $\varphi^*_{\rho^{0} \rho^{0}}$ (top), $\varphi^*_{a_1 \rho^{0}}$ (middle) and  $\varphi^*_{a_1 a_1}$ (bottom), 
for events grouped by the sign of $y_{\rho^{0}}^+ y_{\rho^{0}}^-$, $y_{a_1}^{+} y_{\rho^{0}}^-$ and $y_{a_1}^{+} y_{a_1}^{-}$.
\label{fig:a1a1} }
\end{figure}

Fig.~\ref{fig:a1a1} shows overlaid shapes of the acoplanarity angles distributions for 
scalar (black) and pseudo-scalar (red) Higgs, for positive and 
negative products of   $ y_{\rho^{0}}^{+} y_{\rho^{0}}^{-}$, $ y_{a_1}^{+} y_{\rho^{0}}^{-}$, $y_{\rho^{0}}^{+} y_{a_1}^{-} $, $ y_{a_1}^{+} y_{a_1}^{-}$, given by formulae
(\ref{fig:rhorho}) and (\ref{Eq:ya1}). 
The amplitude of modulation is smaller than in both previously discussed cases, as the 
information on the CP state is now dissipated in the cascade decays of both $\tau$ leptons. 
Also it is smeared due to two possible pairing  of the  $\pi^{+} \pi^{-}$ to form 
$\rho^{0}$ resonance. On the other hand, 16 correlated 1-dimension distributions are available.

\subsection{Case of CP mixing}\label{sec:CPmix}

So far, we have compared distributions of the acoplanarity angles for pure scalar and pseudo-scalar CP states. 
For the mixed scalar-pseudo-scalar case, the general Higgs boson Yukawa coupling to the $\tau$ lepton
\begin{equation}
\bar{\tau} (a + i b \gamma_5)\tau,
\end{equation}
 can be reparametrized with scalar-pseudo-scalar mixing angle $\phi^{CP}$, as   in~\cite{Desch:2003rw}
\begin{equation}
\frac{1}{\cal N} \bar{\tau} (\cos \phi^{CP} + i\ \sin \phi^{CP} \ \gamma_5)\tau.
\end{equation}

\begin{figure}
  \begin{center}                               
{
   \includegraphics[width=7.5cm,angle=0]{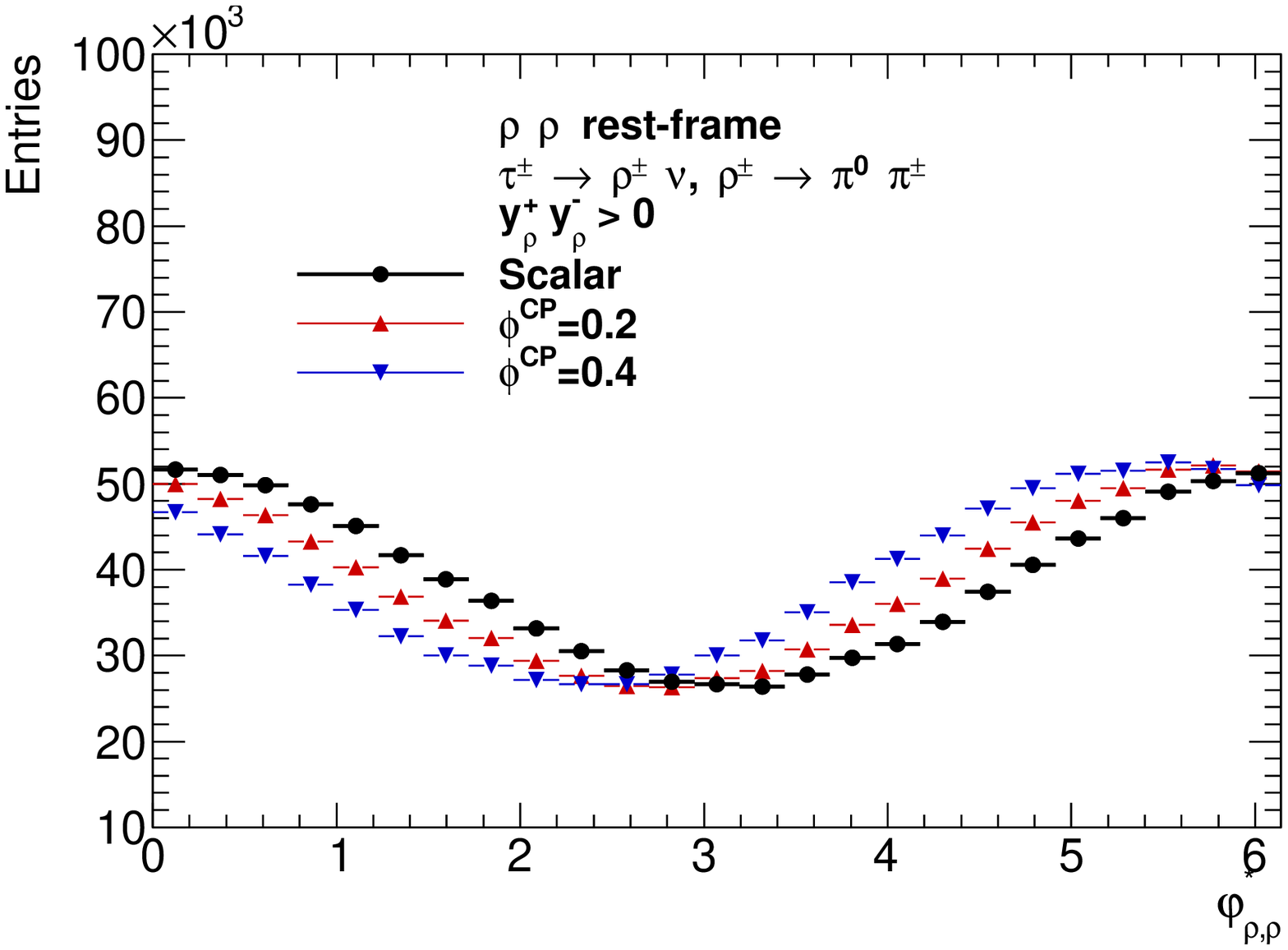}
   \includegraphics[width=7.5cm,angle=0]{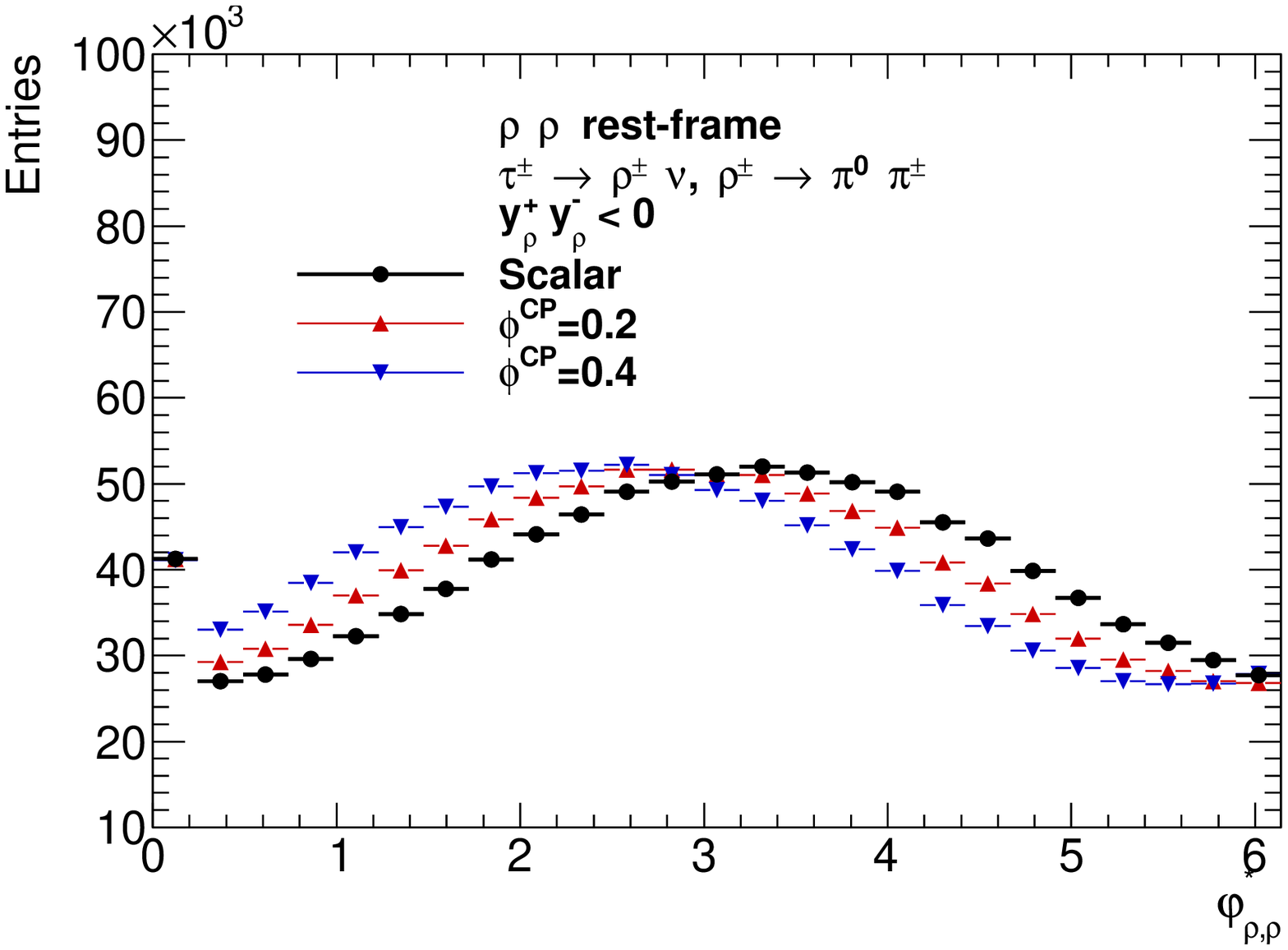}
}
\end{center}
\caption{  Acoplanarity angles of oriented half decay plane, $\varphi^*_{\rho \rho}$, 
for events grouped by the sign of  $y_{\rho}^+ y_{\rho}^-$. Shown are distributions
for three values of mixing angle  $\alpha_{CP}$ = 0.0 (scalar), 0.2 and 0.4.
\label{fig:rhorho_CPmix} }
  \begin{center}                               
{
   \includegraphics[width=7.5cm,angle=0]{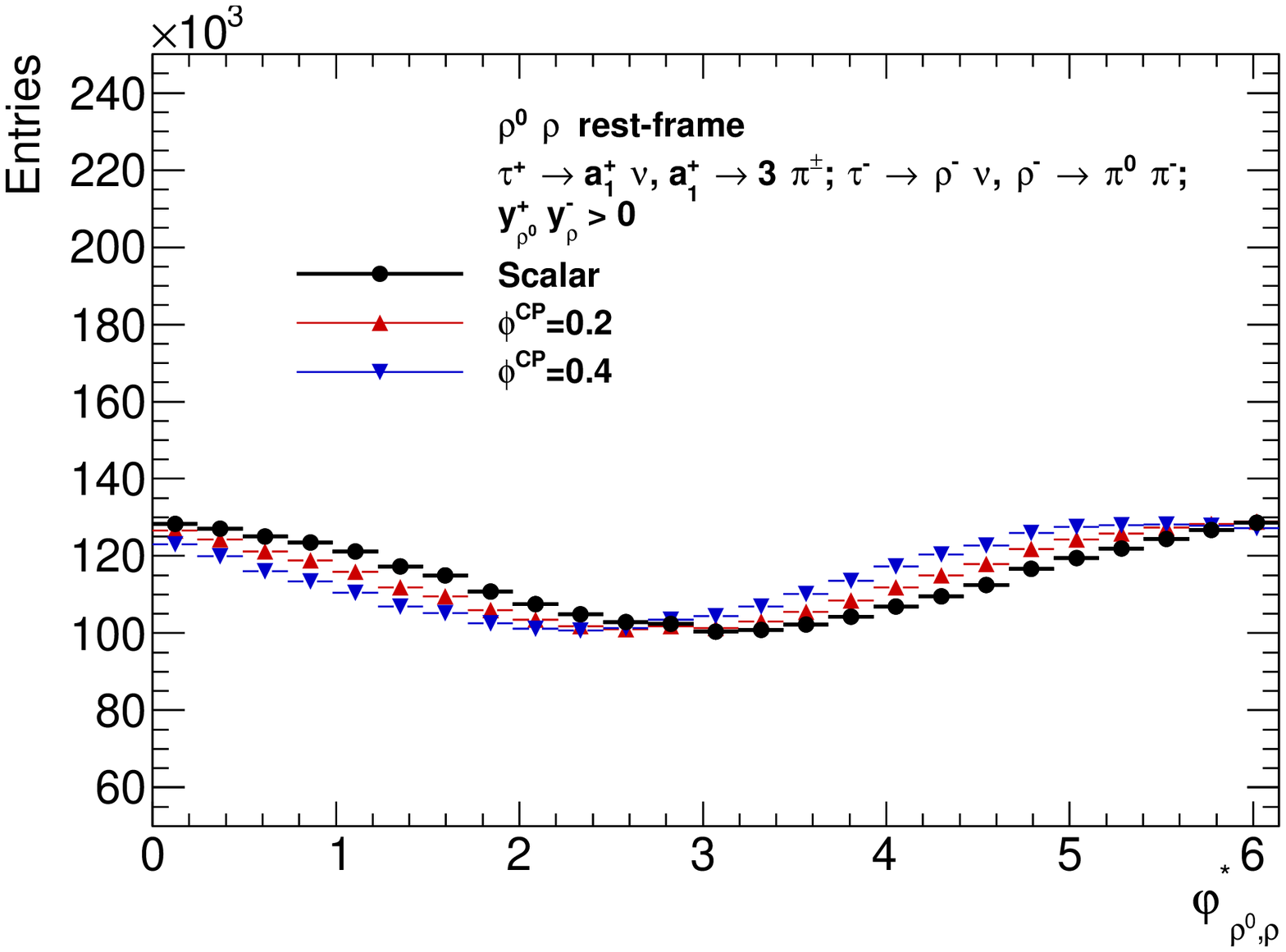}
   \includegraphics[width=7.5cm,angle=0]{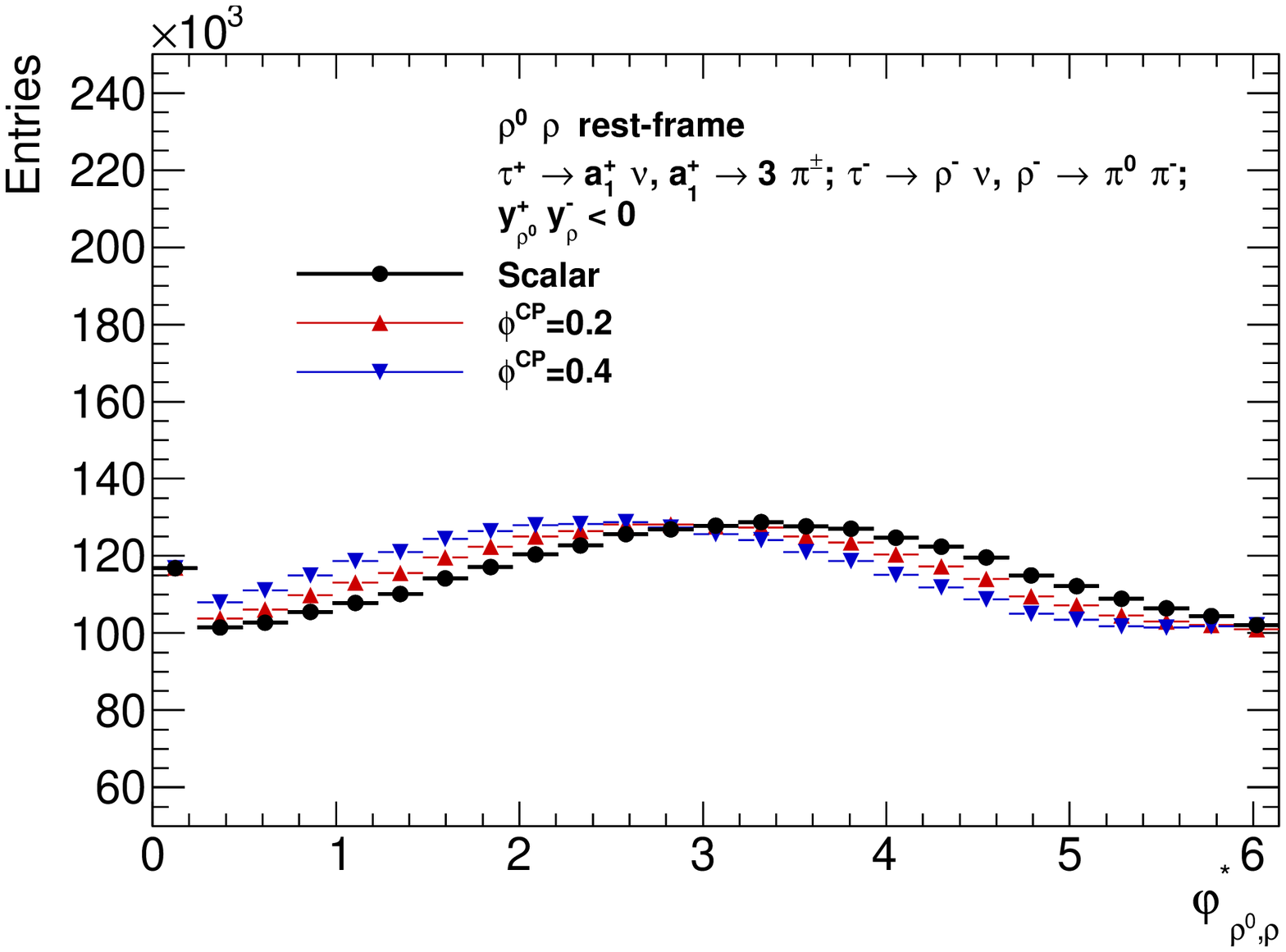}
   \includegraphics[width=7.5cm,angle=0]{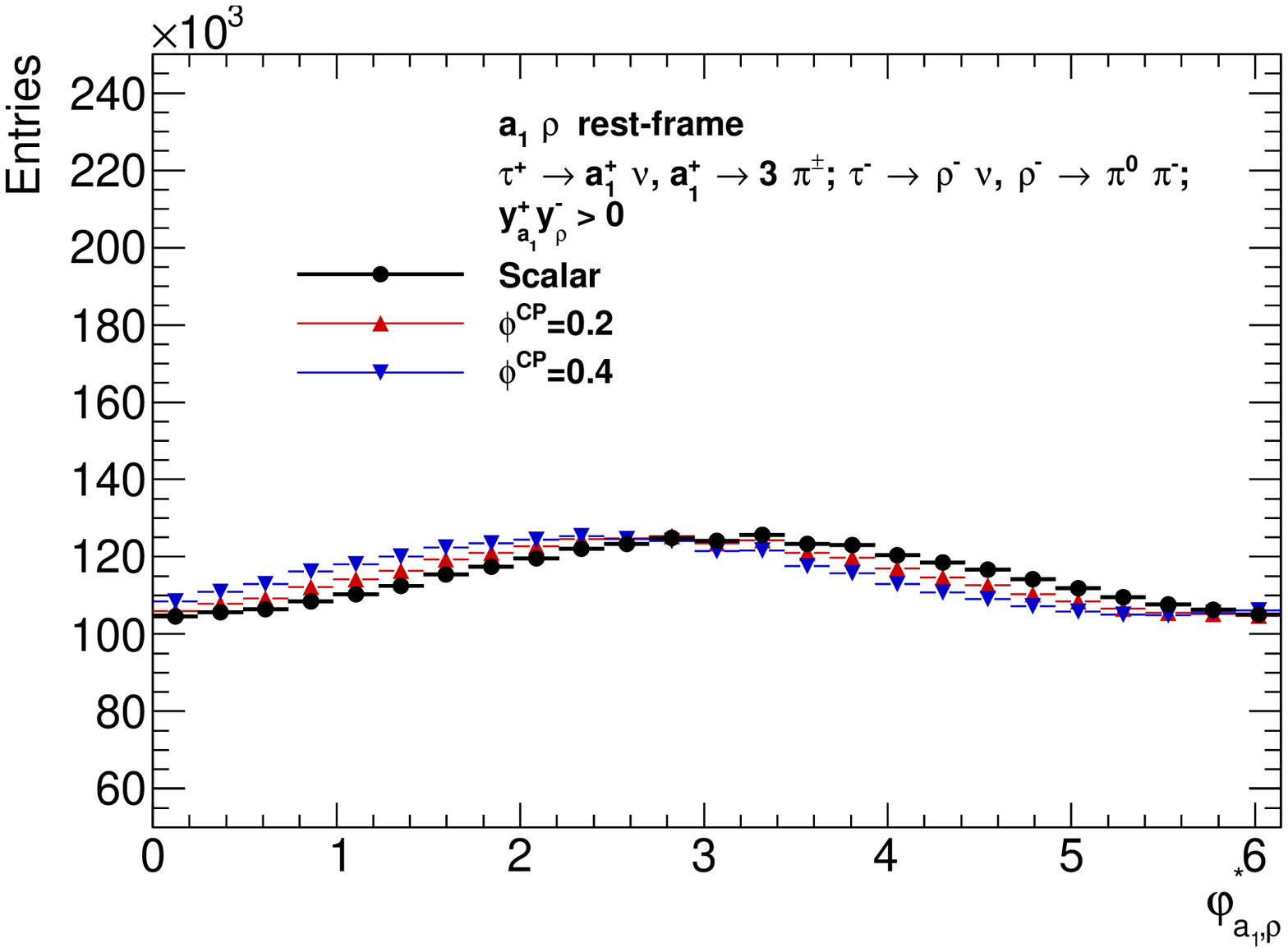}
   \includegraphics[width=7.5cm,angle=0]{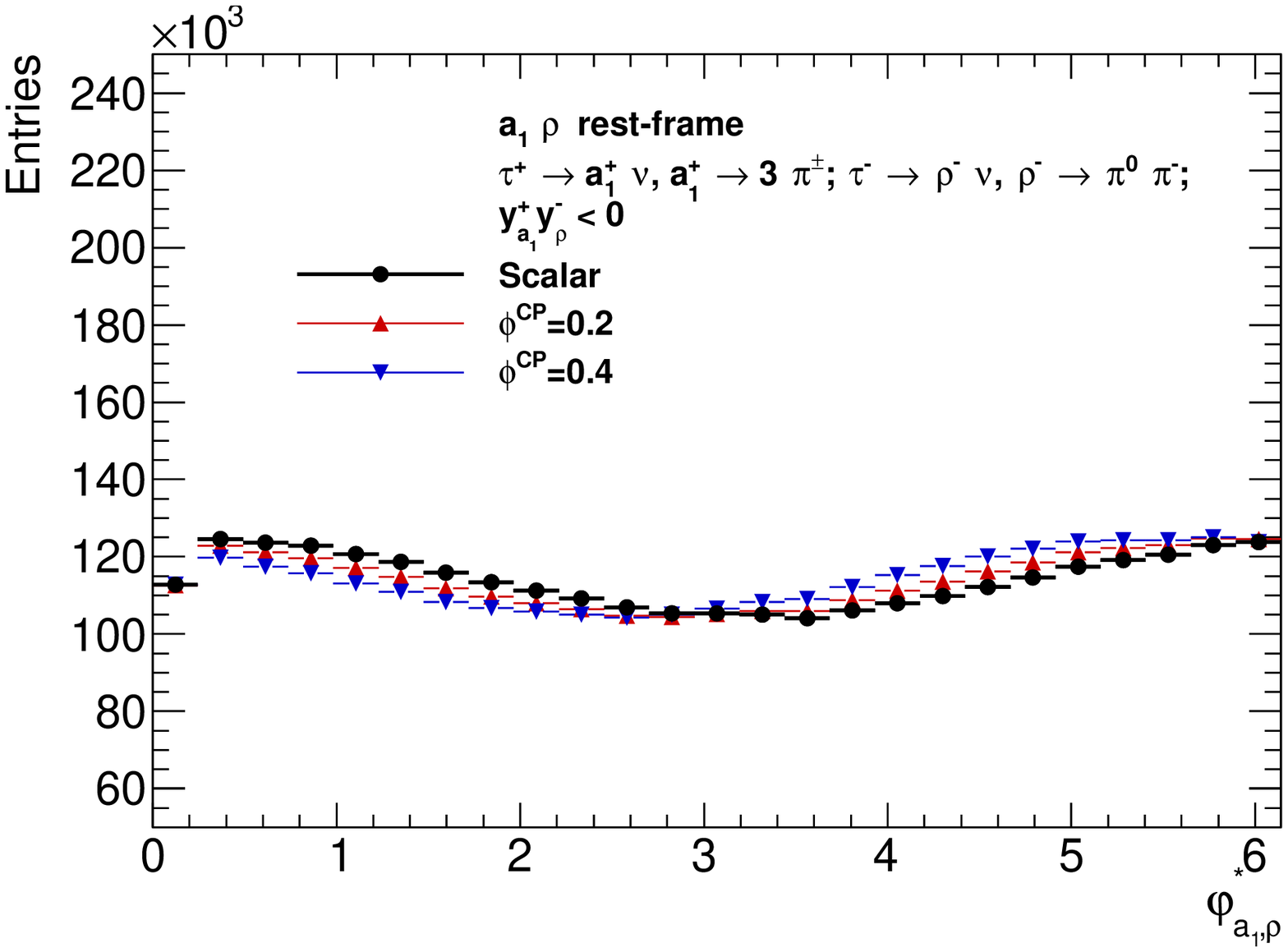}
}
\end{center}
\caption{Acoplanarity angles of oriented half decay planes: $\varphi^*_{\rho^{0} \rho}$ (top), $\varphi^*_{a_1 \rho}$ (bottom), 
for events grouped by the sign of  $y_{\rho^{0}}^+ y_{\rho}^-$ and $y_{a_1}^{+} y_{\rho}^-$ respectively.
 Shown are distributions
for three values of mixing angle  $\phi^{CP}$ = 0.0 (scalar), 0.2 and 0.4.
\label{fig:a1rho_CPmix} }
\end{figure}

\begin{figure}
  \begin{center}                               
{
   \includegraphics[width=7.5cm,angle=0]{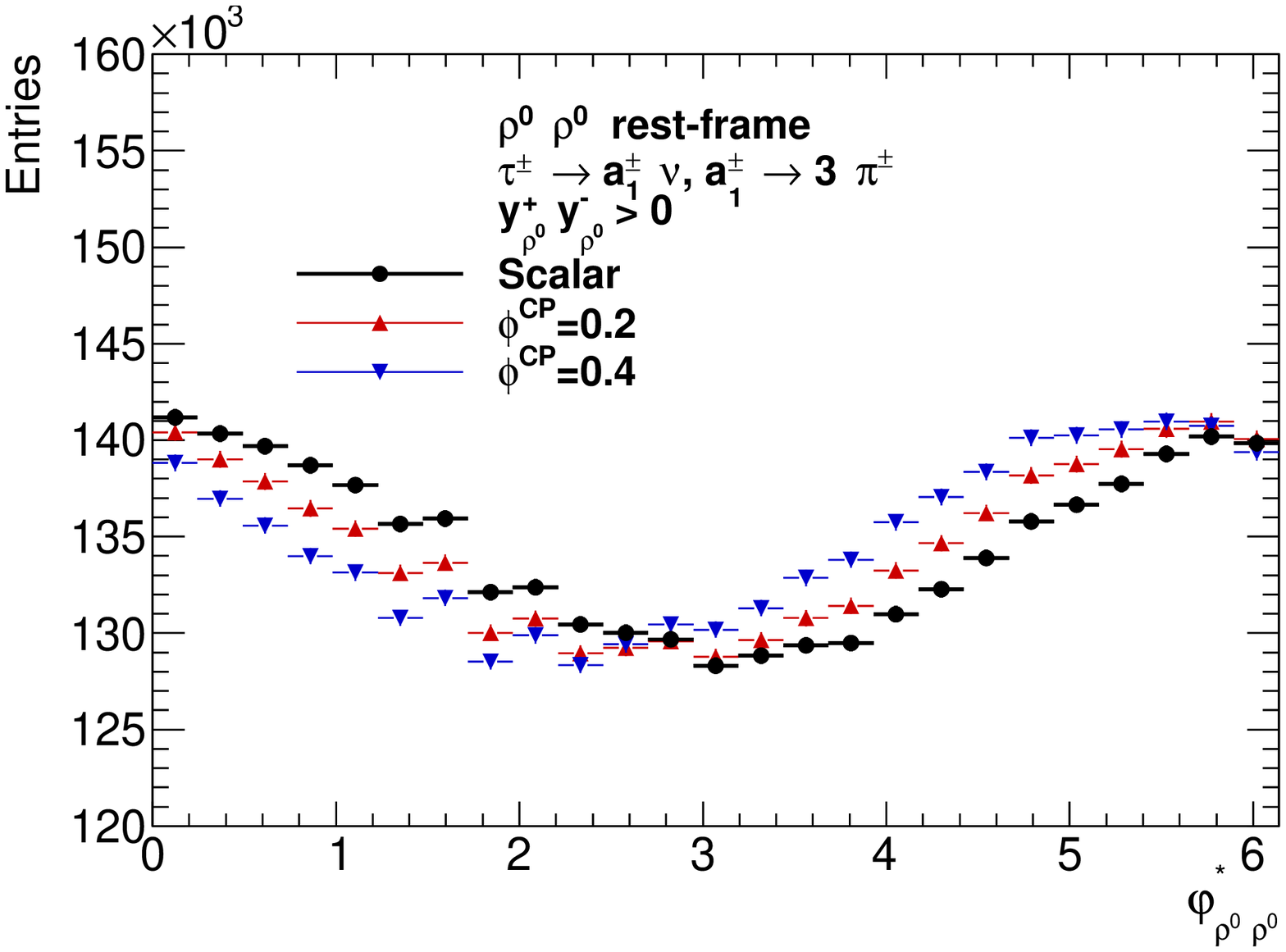}
   \includegraphics[width=7.5cm,angle=0]{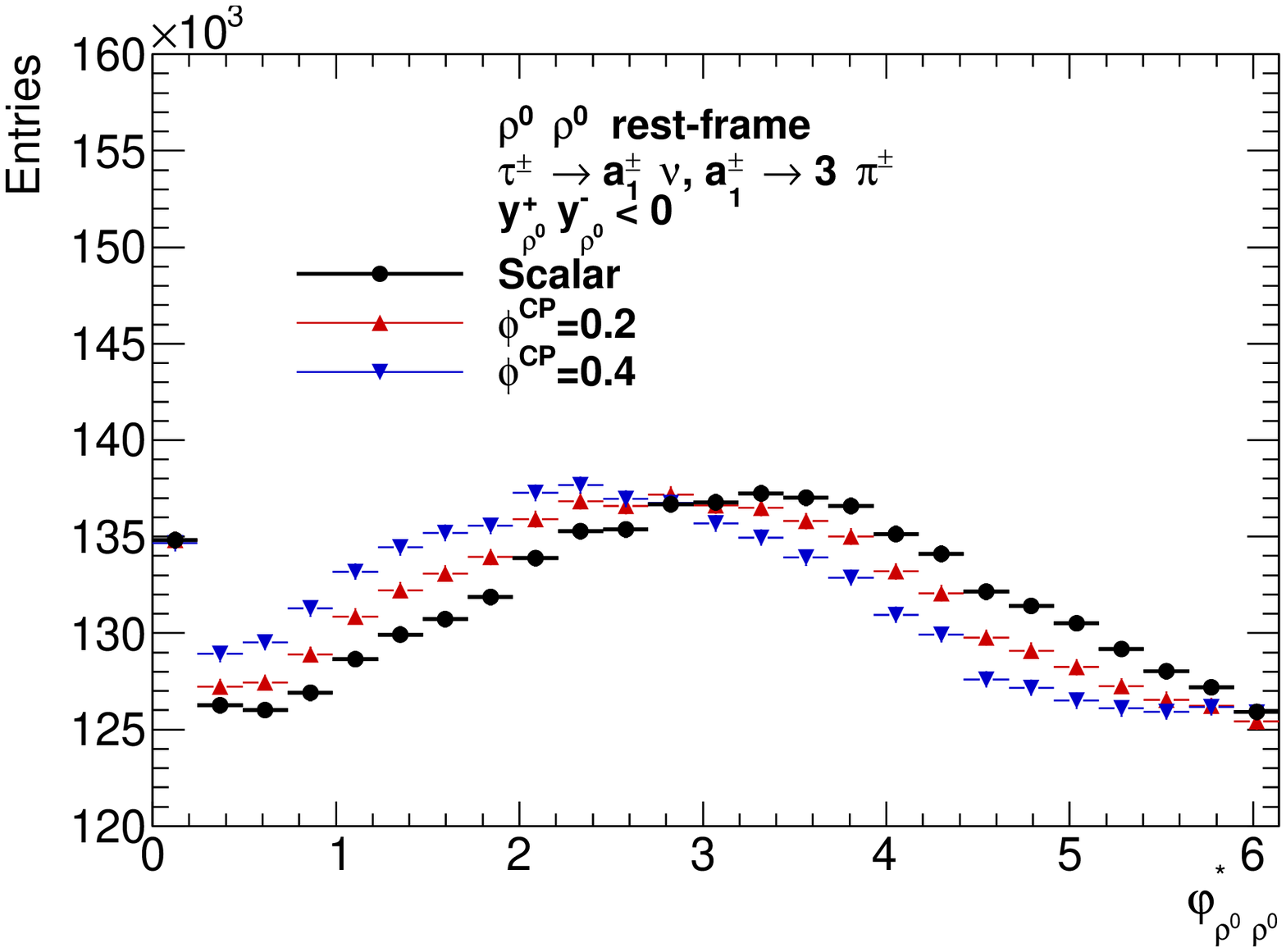}
   \includegraphics[width=7.5cm,angle=0]{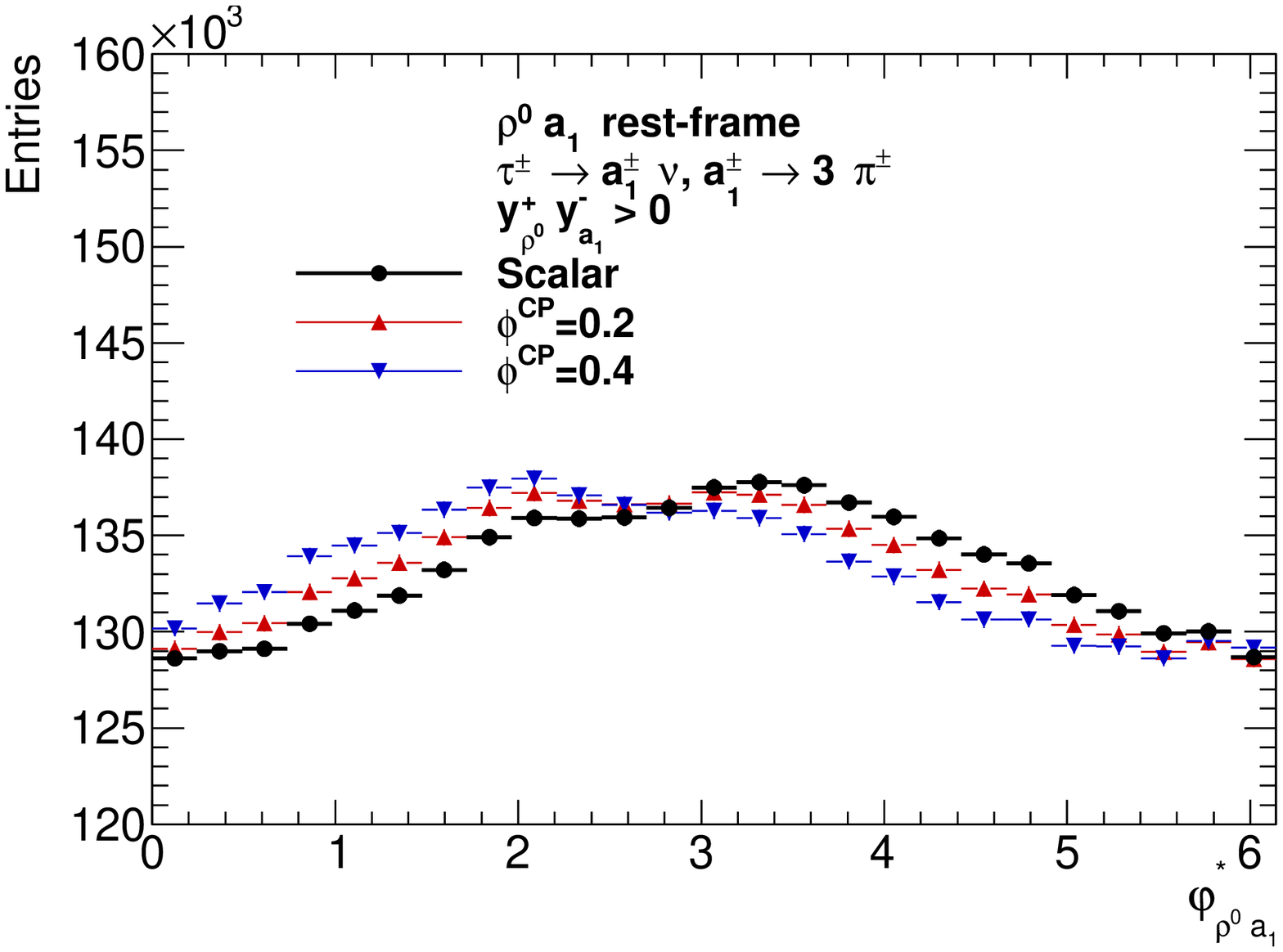}
   \includegraphics[width=7.5cm,angle=0]{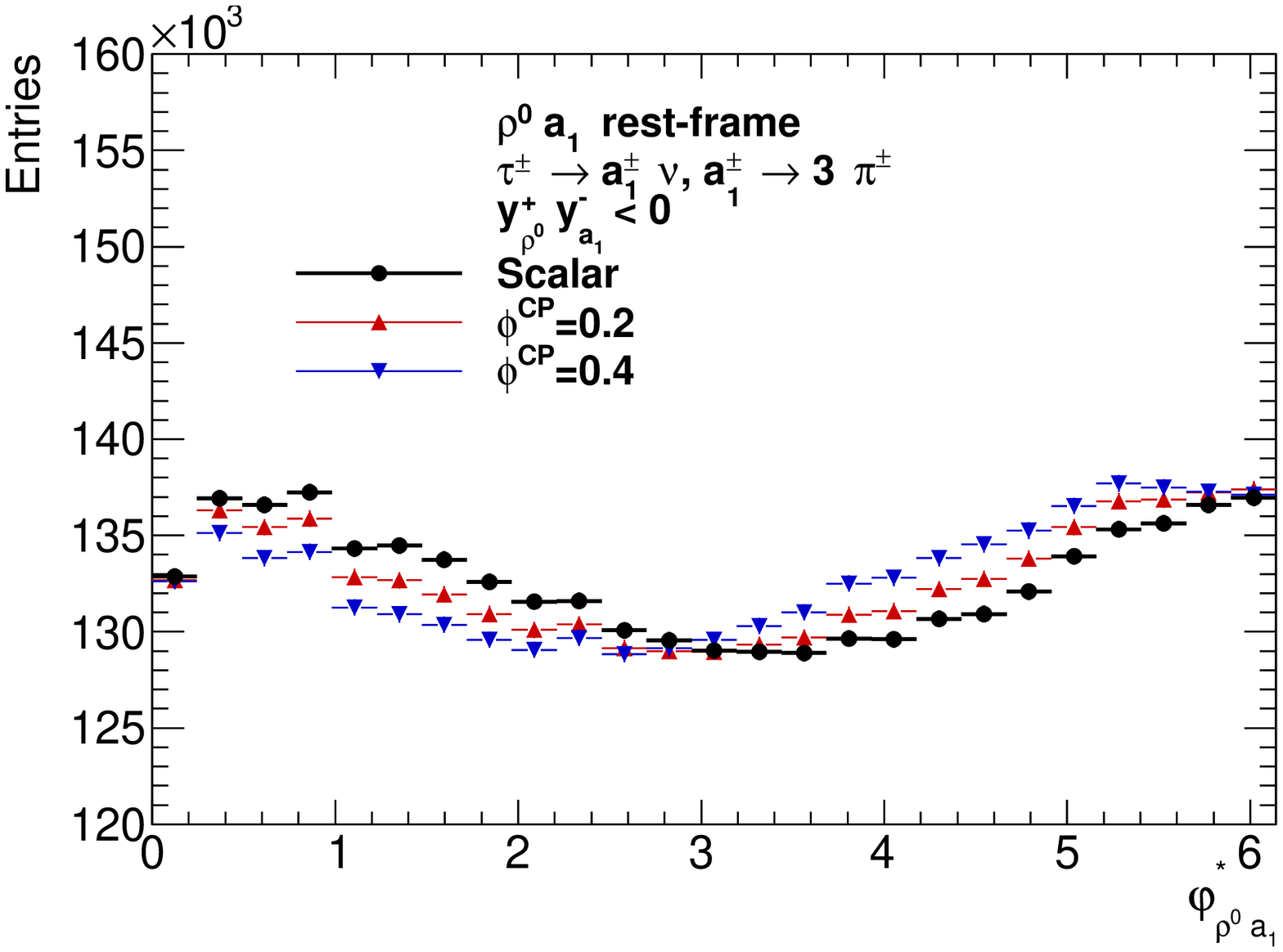}
   \includegraphics[width=7.5cm,angle=0]{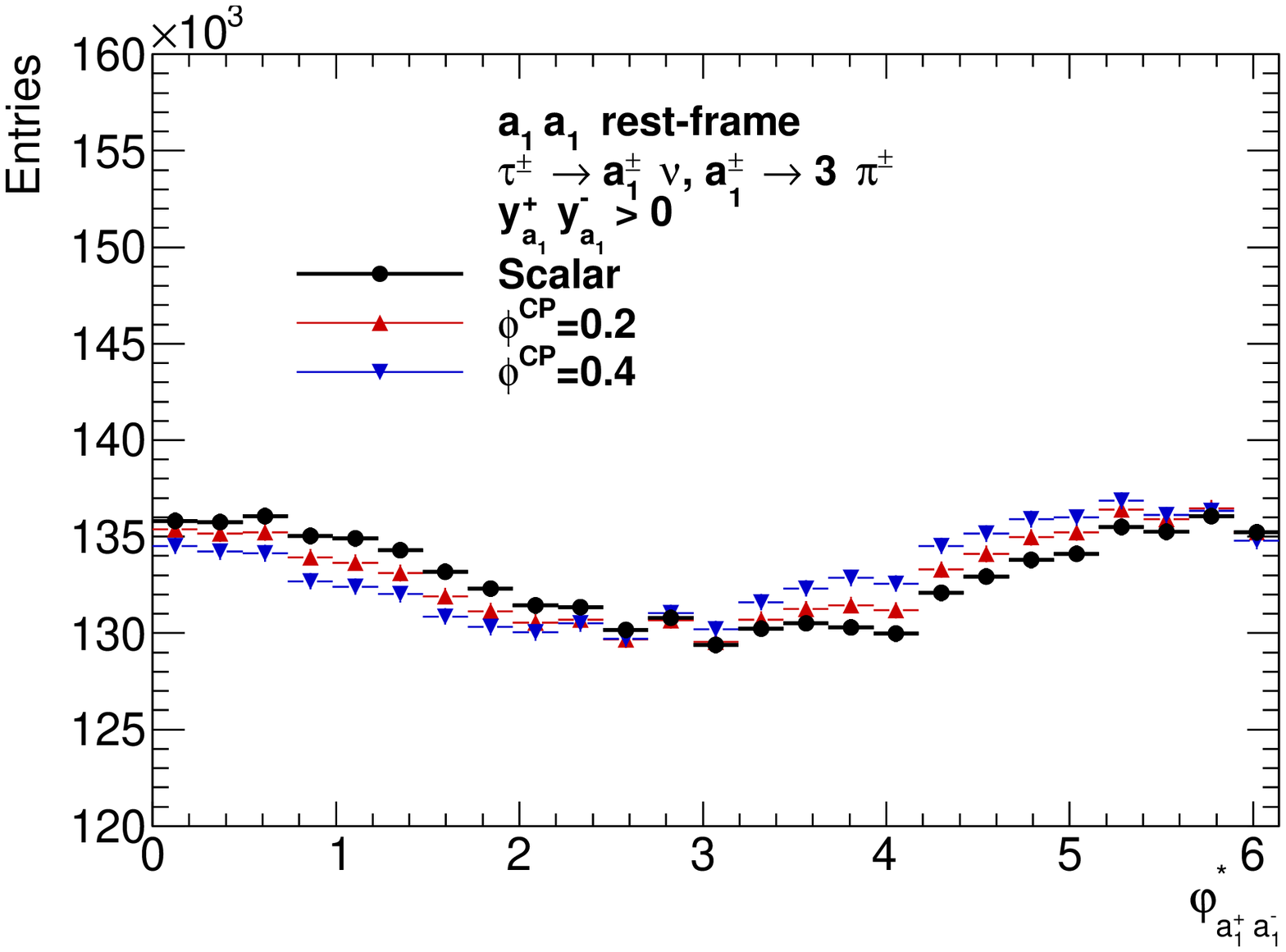}
   \includegraphics[width=7.5cm,angle=0]{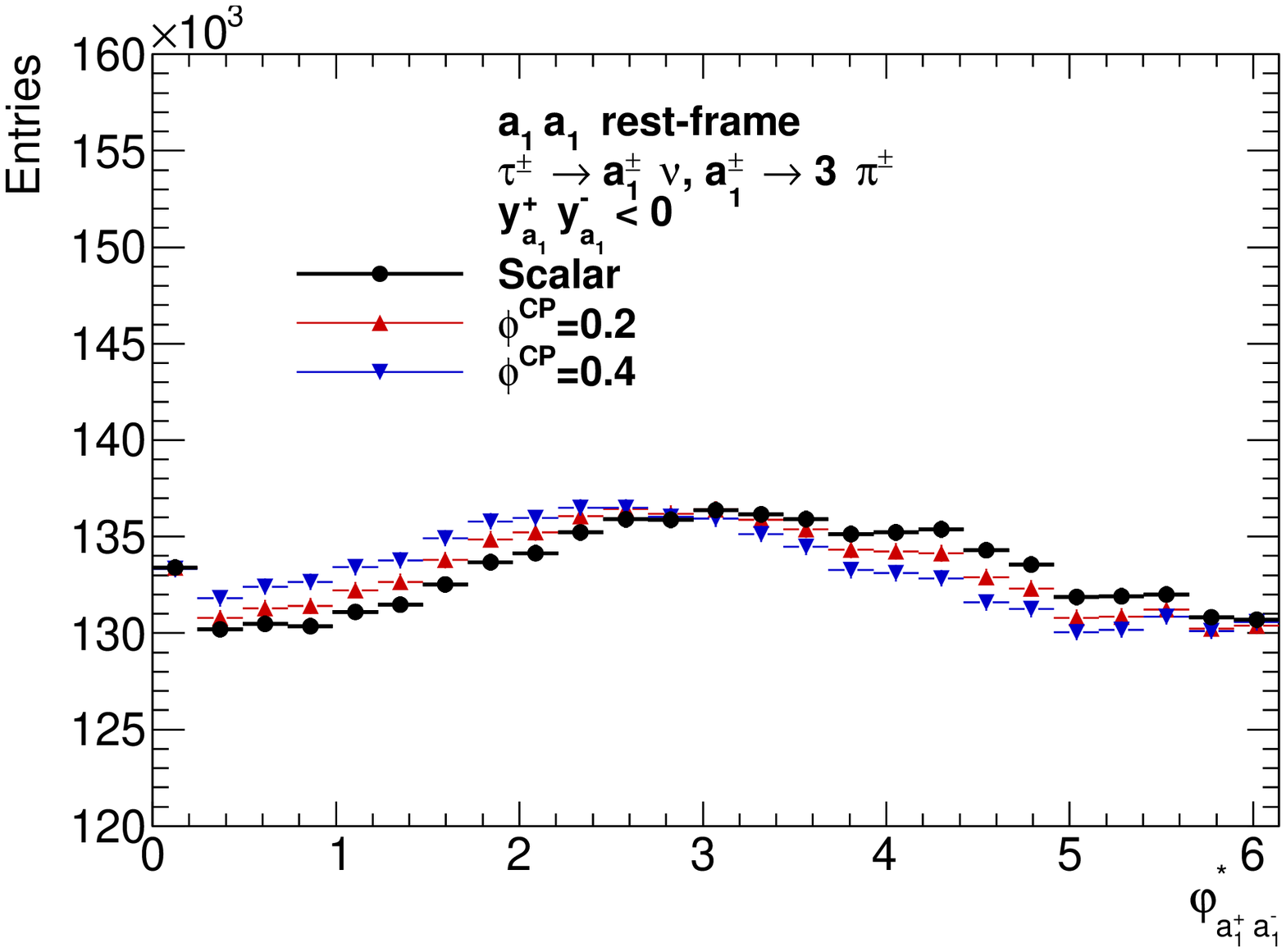}
}
\end{center}
\caption{Acoplanarity angles of oriented half decay planes: $\varphi^*_{\rho^{0} \rho^{0}}$ (top), $\varphi^*_{a_1 \rho^{0}}$ (middle) and  $\varphi^*_{a_1 a_1}$ (bottom), 
for events grouped by the sign of  $y_{\rho^{0}}^+ y_{\rho^{0}}^-$, $y_{a_1}^{+} y_{\rho^{0}}^-$ and $y_{a_1}^{+} y_{a_1}^{-}$ respectively. Shown are distributions
for three values of mixing angle  $\phi^{CP}$ = 0.0 (scalar), 0.2 and 0.4.
\label{fig:a1a1_CPmix} }
\end{figure}

The mixed parity state is simulated by corresponding weight 
$wt^{CP}$ calculated with  {\tt TauSpinner} package~\cite{Przedzinski:2014pla}.
Figures~\ref{fig:rhorho_CPmix} - ~\ref{fig:a1a1_CPmix} show  shift 
in the acoplanarity distributions for cases of
mixing angle $\phi^{CP}$ = 0.0, 0.2 and 0.4 
and three configurations of the $\tau$'s decay modes discussed above.  
Shift in all distributions of acoplanarity angle of 
$\Delta \varphi^* = 2 \phi^{CP}$ is observed, as presented in~\cite{Desch:2003rw} already.

\section{Exploring sensitivity with  ML techniques} \label{sec:NNdeeplearn}

In the discussed so far {\it  1-dimensional analysis}, in each of the decay modes it was possible
to define one or more acoplanarity angles between planes spanned on the observable  $\tau$ leptons decay products. This
was demonstrating sensitivity to the CP state of decaying Higgs boson. 
However, the amplitude in modulation 
of the  acoplanarity angles  distributions,
due to CP parity effect,
was smaller in case of cascade decays like 
$\tau \to a_1 \nu \to \rho^0 \pi^{\pm}\nu \to 3 \pi^{\pm} \nu$ than in 
$\tau \to \rho \nu \to \pi^{\pm} \pi^{0} \nu$. 

With one-dimensional projections of the full phase-space of the $\tau$ decay products, 
some fraction of sensitivity to the CP states could have been averaged out and lost. 
To quantify this effect we have explored the modern technique of the 
Deep Learning Neural Network classification.  The same sample of events as 
presented in Section~\ref{sec:physics}, and the same selection on the visible 
transverse momenta  are used.

The following features (observables/variables) are prepared for 
the Neural Network classification algorithms:
(i) $\varphi^*_{i,k}$ acoplanarity angle or angles\footnote{Defined in the rest-frame of the appropriate resonance 
($a_1, \rho, \rho^0$) pairs.},
(ii) the 4-momenta of the visible
$\tau$ decay products and intermediate resonances 
(iii) virtualities of intermediate resonances 
and (iv) $y_i$ variables i.e. energy differences. 

To eliminate possible  confusion of the Neural Network due to trivial symmetries, we always first boost and rotate event 
to the rest-frame of the sum of visible decay products, where primary resonances  
($ a_{1}^{\pm}$ or $\rho^{\pm}$) are aligned along the z-axis. 
We then investigate if on top of 4-momenta of the $\tau$ decay products, 
the  $\varphi^*_{i,k}$ angles, virtualities of intermediate resonances and 
$y^\pm$ variables offer interesting additional variables for {\tt NN} classification.

Let us now describe the employed ML technique \cite{lecun2015deep} in the more formal and mathematical manner. 

\subsection{Data}
Each data point consists of numerical features, saturated with observables/variables of consecutive events:  the 4-momenta, listed later functions of these 4-momenta, and other information
that detector can capture. We consider three separate problems for $H\to \tau\tau$:
(i) $\tau \tau \to \rho \nu \rho  \nu$,
(ii) $\tau \tau \to a_1 \nu  \rho \nu$,
(iii) $\tau \tau \to a_1 \nu a_1   \nu$.
Depending on the decay modes
of the outgoing $\tau$ pairs, data point requires  different number 
of dimensions to describe. The data point, is called  an event: of the Higgs boson production and  decay 
into $\tau$ lepton pair. The structure of the event can be represented  as follows:

\begin{equation}
  x_{i} = ( f_{i,1}, ..., f_{i, D}), w_{a_{i}}, w_{b_{i}}
\end{equation}

Values $f_{1..D}$ describe features  and $w_a, w_b$ are weights proportional to the
likelihoods that an event comes from set {\tt A} or {\tt B}. They are highly overlapping so the perfect
separation is not possible and $w_a / (w_a + w_b)$ corresponds to the Bayes optimal probability that an event
is sampled from set {\tt A} and not {\tt B}. The $w_{a_i}$ and $w_{b_i}$ are used to compute targets during
the training procedure.

We will solve all three problems using the same neural network architecture that will be described in the following
section. Each event is prepared with a procedure described earlier, and following features are available:
\begin{itemize}
\item
Invariant masses of intermediate resonances:  $m_{i}, m_{k}$ 
\item
Acoplanarity angles $\varphi^*_{i,k}$. The oriented half-planes, of intermediate 
resonances decays, indexed respectively  $i, k$ are used.
\item
Variables $y^+_{i} (y^-_{k})$: visible energy fraction, carried by the first
product of $i$  (or $k$) resonance decay, minus energy fraction, carried by the
second product.  Mass corrections of formula 
(\ref{Eq:ya1}) applied if necessary. 
\item
The 4-momenta of outgoing visible decay products and intermediate resonances.
In case of cascade decays, we provide 4-momenta of all $\pi^+\pi^-$ pairs which can
form the resonances.
\end{itemize}
For the configuration (i) only one pairing of outgoing $\pi^0, \pi^{\pm}$  into intermediate resonance is possible, 
for configuration (ii) there are two pairings of $\pi^\pm, \pi^{\mp},$ and for configuration (iii) there are 4 pairing.
Thanks to the fact that model {\tt A} and {\tt B} are prepared using the same sample of events, but only with different spin 
weight $wt^{CP}$, the statistical fluctuations are largely reduced. 
It has also consequence on the actual implementation of the ML metrics and code.  
For the numerical studies discussed below, we will use subset or full set of 
the above event features,
see Table~\ref{tab:DeepLearn}.

Afterward each feature column is centered to have 0 mean and 1 standard deviation across the training data-set.

\subsection{Metric}
We need a quantitative metric to compare different models and approaches. The metric used to compare the models
is a weighted Area Under Curve {\tt (AUC)} \cite{AUC}. $X = (x_1,..., x_n)$ is a dataset of interest consisting of $n$ events.
$p = (p_1, .., p_n)$ is a vector probabilities returned by neural network model,
$p_i = p(x_i \in A) = 1 - p(x_i \in B)$.

The final metric will be computed as follows:
\begin{eqnarray}
SCORE(X, p) &=& weightedAUC(  \nonumber \\
    & & [(1, p_1, w_{a_1}), (0, p_1, w_{b_1}), \nonumber \\
    & &  (1, p_2, w_{a_2}), (0, p_2, w_{b_2}), \\
    & & ...                             \nonumber   \\ 
    & &  (1, p_n, w_{a_n}), (0, p_n, w_{b_n})])\nonumber 
\end{eqnarray}

That is, each event contributes twice to the computation of the score:
\begin{itemize}
\item
  $(1, p_i, w_{a_i})$ - corresponds to the case in which model correctly predicts $x_i$ in {\tt A} with assigned
  probability $p_i$.  It contributes to the final loss with weight $w_{a_i}$.
\item
  $(0, p_i, w_{b_i})$ - corresponds to the case in which model incorrectly predicts $x_i$  in {\tt A} with assigned
  probability $p_i$. It contributes to the final loss with weight $w_{b_i}$.
\end{itemize}
The first value in a tuple represents the true target (1 means $x_i$ in A), the second is used for ranking events
for the purposes of {\tt AUC} and the last value represents the weight associated with the event.

$SCORE(X, p)$ will return a value of 0.5 for a model that assigns random predictions. Score of 1.0 is reached for
perfect separation of the distributions. In practice, perfect score is not achievable on these problems as
the distributions are overlapping. It can be shown that the best achievable score is reached when using
optimal predictions $(p_i = w_{a_i} / (w_{a_i} + w_{b_i}))$, which corresponds to about 0.782 result (slightly varies
by problem). We will call them oracle predictions.

\subsection{Model}

Deep Neural Networks~\cite{lecun2015deep} have been shown to work very well across many different domains, including image classification,
machine translation or speech recognition. We will apply similar techniques
to our problem. Neural network can be
seen as a non-linear map between inputs and outputs. They are often build using chains of matrix multiplications
separated by element-wise transformations. 
We want to distinguish between the two different CP states  of Higgs particle so we frame the problems as binary
classifications.

The basic architecture used for the problems will contain D-dimensional input (problem dependent) followed
by matrix multiplications transforming the input into N-dimensional space with a ReLU non-linearity
$(ReLU(x) = max(0, x))$. We apply multiply such transformations to add more expressive power to the model.
The largest network used in our experiments had 7~matrix multiplications transforming data points into following
sizes 
\begin{equation}
D\rightarrow 300\rightarrow 300\rightarrow 300\rightarrow 300\rightarrow 300\rightarrow 300\rightarrow 1 \nonumber
\end{equation}
The output is a scalar value that represents an indicator
on whether an event looks closer to type {\tt A} or {\tt B}. We would like to represent the output of
a neural network as a probability between the two choices. A common way to accomplish this, and also used here,
is to use sigmoid non-linearity $(sigmoid(x) = 1 / (1 + exp(-x)))$ on the last layer, which squishes the output
into interval $[0.,1]$ and can be interpreted as probability. The metric minimized by the model is negative
log likelihood of the true targets under Bernoulli distributions. That corresponds to a loss function equal to:
\begin{equation}
  - log p(y | y_h)  =  - (y == 0) * log(y_h) - (y == 1) * log(1 - y_h),
  \end{equation}
where $y_h$ represents probability outputted by neural network model.

Initially the weights of the matrices are picked randomly and are optimized using a variant of stochastic
gradient descent algorithm called Adam~\cite{kingma2014adam}. We also used a recent trick called
Batch Normalization~\cite{ioffe2015batch} and Dropout~\cite{srivastava2014dropout} to improve the training of neural network
model. Everything was implemented using {\tt TensorFlow}~\cite{abadi2015tensorflow}, an open-source framework 
for numerical computations.

The snippet of the code with Neural Network model, in {\tt python} programming language, prepared for this analysis, 
is included in Appendix A. 

\subsection{Results}
In Figs \ref{fig:a1rho}, \ref{fig:a1a1} and also \ref{fig:a1rho_CPmix}, \ref{fig:a1a1_CPmix} 
of Section \ref{sec:physics}, 
we have demonstrated, that in the case of 
one-dimensional histograms and $\tau \to a_1 \nu$ decays, several variants of acoplanarity 
angles and CP sensitive observables can be defined. However, question of estimation of the overall
statistical significance was difficult.
Now, in the analysis of the {\tt NN} model we have correlated information on 
the acoplanarity angles and 
signs of energy differences,
with information on the invariant masses of the $\rho^{\pm}$, $\rho^{0}$ and $a_1^{\pm}$ resonances in the decay 
and individual directions of the final decay products in an automated way.
We could also profit of the automated control of the correlations between distributions of
the distinct acoplanarity angles. 
Investigation of  improvement for the discriminating 
power, if some of the variables such as $\varphi_{i,k}^*$, $y_i^\pm$, $y_k^\mp$, $m_i$, $m_k$ 
or 4-momenta of the outgoing visible 
decay products of the $\tau^\pm$ leptons were used or not, was possible. 

\begin{table}
 \vspace{2mm}
  \begin{center}
  \begin{tabular}{|l|r|r|r|r|}
  \hline\hline 
  Features/variables      & Decay mode: $\rho^{\pm}- \rho^{\mp}$    &  Decay mode: $a_1^{\pm} - \rho^{\mp}$   & Decay mode:  $a_1^{\pm} - a_1^{\mp}$  \\ 
                          & $\rho^{\pm} \to \pi^{0}\ \pi^{\pm}$      &  $ a_{1}^{\pm} \to \rho^{0} \pi^{\pm},\ \rho^{0} \to  \pi^{+} \pi^{-}$  
                                                                  &  $ a_{1}^{\pm} \to \rho^{0} \pi^{\pm},\ \rho^{0} \to  \pi^{+} \pi^{-}$  \\ 
                          &                                       &  $\rho^{\mp} \to \pi^{0}\ \pi^{\mp}$   &                                     \\
  \hline\hline
  \hline
  $\varphi^*_{i,k}$                                           & 1      &   4          &  16    \\
  \hline
  $\varphi^*_{i,k}$ and $y_i, y_k$                            & 3      &   9          &  24    \\ 
  \hline
  $\varphi^*_{i,k}$, 4-vectors                                & 25     &  36          &  64    \\ 
  \hline
  $\varphi^*_{i,k}$, $y_i, y_k$ and $m_i, m_k$                 & 5      &  13          &  30    \\ 
  \hline
  $\varphi_{i,k}^*$, $y_i$, $y_k$, $m_i$, $m_k$ and 4-vectors  & 29     &  45          &  78    \\ 
\hline
\end{tabular}
\end{center}
\caption{Dimensionality  of the features which may be used in each discussed configuration of the decay modes. 
Note that 
in principle  $y_i^\pm$, $y_k^\mp$ may be calculated in the rest frame of the resonance pair used to define $\varphi_{i,k}^*$ planes,
but in practice, choice of the frames is of no numerically significant effect. We do not distinguish such variants.
 } 
\label{tab:DeepLearnDim}
\end{table}

Table~\ref{tab:DeepLearnDim} summarizes dimensions of each variant for the input considered.
Results of the performance with {\tt NN} model implemented using {\tt TensorFlow} framework are summarized
in Tables~\ref{tab:DeepLearn}~and~\ref{tab:DeepLearn_CPmix}. Primary interest was to access and quantify how much 
information is still available in the correlations between 4-vectors of outgoing $\tau$-leptons decay products
which is not captured in the 1-dimensional projections on the $\varphi^*$ angles. Of interest was also to which extend {\tt NN} model
can capture non-trivial correlations (like the $\varphi^*$ angles) given only simple information on 4-vectors of outgoing particles.

We have found, that minimal requirement was to boost all visible decay 
products into rest-frame of all  visible final decay 
products of Higgs. Alignment along $z$ axis of combined  visible decay products 
for  $\tau^+$ and  $\tau^-$ was also necessary.
As expected, the use of   $\varphi_{i,j}^*$ alone was not sufficient to provide any discrimination. 
On the other hand, if only 4-vectors of visible final state scalars were provided,
classificator could distinguish between scalar and pseudo-scalar
models.
Further, higher level  variables, such as  $\varphi_{i,j}^*$,  $y_i^\pm$ 
or $m_i$ were improving performance only slightly.

When we were reducing the  set of these variables, see Tables
\ref{tab:DeepLearn}~and~\ref{tab:DeepLearn_CPmix}, efficiency of 
the {\tt NN} deteriorated.
This  provides interesting insight into performance of {\tt NN}.
However its practical consequences may  escape 
conclusive interpretation because  
detector effects are not taken into account.

\begin{table}
 \vspace{2mm}
  \begin{center}
  \begin{tabular}{|l|r|r|r|r|}
  \hline\hline 
  Features/var-           & Decay mode: $\rho^{\pm}- \rho^{\mp}$    &  Decay mode: $a_1^{\pm} - \rho^{\mp}$   & Decay mode:  $a_1^{\pm} - a_1^{\mp}$  \\ 
  iables                  & $\rho^{\pm} \to \pi^{0}\ \pi^{\pm}$   &  $ a_{1}^{\pm} \to \rho^{0} \pi^{\mp},\ \rho^{0} \to  \pi^{+} \pi^{-}$  
                                                               &  $ a_{1}^{\pm} \to \rho^{0} \pi^{\pm},\ \rho^{0} \to  \pi^{+} \pi^{-}$             \\ 
                          &                                   &  $\rho^{\mp} \to \pi^{0}\ \pi^{\mp}$   &                                         \\
  \hline\hline
  True classification                             & 0.782       &  0.782          &  0.782    \\
  \hline
  $\varphi^*_{i,k}$                                       & 0.500       &  0.500         &  0.500     \\
  \hline
  $\varphi^*_{i,k}$ and $y_i, y_k$                             & 0.624       &  0.569          &  0.536    \\
  \hline
    4-vectors                                                & 0.638       &  0.590          &  0.557    \\
  \hline
  $\varphi^*_{i,k}$, 4-vectors                            & 0.638       &  0.594          &  0.573    \\
  \hline
  $\varphi^*_{i,k}$, $y_i, y_k$ and $m^2_i, m^2_k$                   & 0.626       &  0.578          &  0.548    \\
  \hline
  $\varphi_{i,k}^*$, $y_i$, $y_k$, $m^2_i$, $m^2_k$ and 4-vectors         & 0.639       &  0.596          &  0.573    \\
\hline
\end{tabular}
\end{center}
\caption{ Average probability $p_i$ that a model predicts correctly event $x_i$ to be of a  type $A$ (scalar),
with training being performed for separation between type $A$ and $B$ (pseudo-scalar). } 
\label{tab:DeepLearn}
\end{table}

\begin{table}
 \vspace{2mm}
  \begin{center}
  \begin{tabular}{|l|r|r|r|r|}
  \hline\hline 
  Features/variables                             & $\phi^{CP}$ = 0.2    &   $\phi^{CP}$ = 0.3   &  $\phi^{CP}$ = 0.4  \\ 
  \hline\hline
  True classification                            & 0.560          & 0.588          & 0.616        \\
  \hline
  $\varphi_{i,k}^*$, $y_i$, $y_k$, $m^2_i$, $m^2_k$ and 4-vectors      & 0.526          & 0.540          & 0.553        \\
\hline
\end{tabular}
\end{center}
\caption{ Average probability $p_i$ that a model predicts correctly event $x_i$ to be of a  type $A$ (scalar),
with training being performed for separation between type $A$ and $B$ (CP-mix state) with mixing angle of 
$\phi^{CP}$ =0.2, 0.3, 0.4 respectively. Results are shown only for  $\rho^{\pm}- \rho^{\mp}$ decay mode.} 
\label{tab:DeepLearn_CPmix}
\end{table}

\section{Summary} \label{sec:summary}

We have demonstrated that for the measurement of  Higgs boson CP properties in $H \to \tau \tau$ channel the 
use of $ \tau^{\pm} \to a_{1}^{\pm} \nu; \;  a_{1}^{\pm} \to 3 \pi^{\pm}$ decay, together with well known  
$ \tau^{\pm} \to\rho^{\pm} \nu;\;  \rho^{\pm} \to \pi^{\pm} \pi^{0}$ decay mode is promising. It  almost double available statistics of signal events:
from 6.5 to 11.9 \% of all $H \to \tau\tau$ decays.

We have shown that 1-dimensional acoplanarity angles sensitive to the CP states of the decaying 
Higgs boson, as already established for  $\tau^{\pm} \to\rho^{\pm} \nu;\;  \rho^{\pm} \to \pi^{\pm} \pi^{0}$ case, can be used for  $a_{1}^{\pm}$ decay channel
as well. There can be however, up to 16 of such angles defined for each event.
This provide an interesting set of variables for the 
{\it 1-dimensional analysis}, which can be combined with advanced statistical methods for fitting
templates and extracting information on the CP states.

We have investigated  sensitivity if  one-dimensional projections of multi-dimensional
phase-space are used only. Then, some correlations between directions of outgoing 
particles (thus acoplanarities) are not exploited. To quantify sensitivity, we have 
developed Deep Learning Neural Network model in the {\tt TensorFlow} framework. We have quantified expected sensitivity and, in particular,
we have shown improvement in the discrimination power coming from the higher level variables.
Discrimination probability    of up to 0.596 per single $\tau\tau \to a_1^\pm \rho^\mp \nu\nu$ event, was achieved. This is not much smaller than 0.639 
of $\tau\tau \to \rho^\pm \rho^\mp \nu\nu$ events.
The respective core part of the model in  {\tt TensorFlow} framework is shown in the Appendix A.

In the presented discussions, we have not used information from the decay vertex. On the other hand we have exploited
properties of cascade decays of $\tau$ into broad resonances; first $a_1$ and later $\rho^0$.
 We have found it  useful. We have not discussed ambiguities due to 
modeling of $\tau \to a_1 \nu$ decays though.  

Evaluation if the method can be used in practice, require careful study
of detector ambiguities and background contaminations. In case when only 
$\tau^\pm \to \pi^\pm \pi^0 \nu $ decay modes were to be used, consequences
of the ambiguities were discussed in the literature and were found to be acceptable.
We expect, 
that in this respect the  $a_1^\pm \to (3\pi)^\pm$  decay case is not 
of essential difference.

The  $ a_{1}^{\pm} \to  \pi^{\pm} \pi^{0}  \pi^{0}$ decay is  more challenging,
because two partly overlapping in detector $\pi^{0}$ have to be resolved.
That is why, for now,  we have omitted it from our study, even if it would 
increase statistics sizably.
 On the other hand,
it will have to be taken into account, at some point anyway, 
because it forms important background for $ \tau^{\pm} \to \pi^{\pm} \pi^{0} \nu$
decay.

\vskip 1 cm
\centerline{\bf \Large Acknowledgments}
\vskip 0.5 cm

ERW and ZW were supported in part from funds of Polish National Science
Centre under decisions UMO-2014/15/ST2/00049 and by the Research Executive 
Agency (REA) of the European Union under the Grant Agreement PITNGA2012316704 (HiggsTools).

Majority of the simulation were performed at  PLGrid Infrastructure of the Academic 
Computer Centre CYFRONET AGH in Krakow, Poland.

\providecommand{\href}[2]{#2}
\clearpage
\appendix
\section{Python code defining Neural Network model}
\label{NeuralNetworkClass}

\begin{verbatim}
# Linearly transforms X of shape [batch_size, size1] into [batch_size, size].
# Applies X -> XW + b, where W and b are trainable parameters.
def linear(x, name, size, bias=True):
    w = tf.get_variable(name + "/W", [x.get_shape()[1], size])
    b = tf.get_variable(name + "/b", [1, size],
                        initializer=tf.zeros_initializer)
    return tf.matmul(x, w) + b


# Applies batch normalization trick from https://arxiv.org/abs/1502.03167
# by normalizing each feature in a batch.
def batch_norm(x, name):
    mean, var = tf.nn.moments(x, [0])
    normalized_x = (x - mean) * tf.rsqrt(var + 1e-8)
    gamma = tf.get_variable(name + "/gamma", [x.get_shape()[-1]],
        initializer=tf.constant_initializer(1.0))
    beta = tf.get_variable(name + "/beta", [x.get_shape()[-1]])
    return gamma * normalized_x + beta


class NeuralNetwork(object):

    def __init__(self, num_features, batch_size, num_layers=6, size=300, lr=1e-3):
        # Each input x is represented by a given number of features
        # and corresponding weights for target distributions A and B.
        self.x = x = tf.placeholder(tf.float32, [batch_size, num_features])
        self.wa = wa = tf.placeholder(tf.float32, [batch_size])
        self.wb = wb = tf.placeholder(tf.float32, [batch_size])
        # The model will predict a single number, which is a probability of input x
        # belonging to class A. That probability is equal to wa / (wa + wb).
        y = wa / (wa + wb)
        y = tf.reshape(y, [-1, 1])

        # We apply multiple layers of transformations where each layer consists of
        # linearly transforming the features, followed by batch normalization (described above)
        # and ReLU nonlinearity (which is an elementwise operation: x -> max(x, 0))
        for i in range(num_layers):
            x = tf.nn.relu(batch_norm(linear(x, "linear_%d" % i, size), "bn_%d" % i))

        # Finally, the output is tranformed into a single number.
        # After applying sigmoid non-linearity (x -> 1 / (1 + exp(-x))) we'll interpret that number 
        # as a probability of x belonging to class A.
        x = linear(x, "regression", 1)
        self.p = tf.nn.sigmoid(x)

        # The objective to optimize is negative log likelihood under Bernoulli distribution:
        # loss = - (p(y==A) * log p(y==A|x) + p(y==B) * log p(y==B|x))
        self.loss = loss = tf.reduce_mean(tf.nn.sigmoid_cross_entropy_with_logits(x, y))
        # The model parameters are optimized using gradient-based Adam optimizer
        # (https://arxiv.org/abs/1412.6980) to minimize the loss on the training data.
          self.train_op = tf.train.AdamOptimizer(lr).minimize(loss)

\end{verbatim}

\end{document}